\documentclass[twocolumn,aps,prl,showpacs,amsmath,superscriptaddress,longbibliography,notitlepage,floatfix]{revtex4-2}
\usepackage{amssymb}
\usepackage{mathrsfs}
\usepackage{graphicx}
\usepackage{float}
\usepackage[caption=false]{subfig}
\usepackage[pdftex,colorlinks=red]{hyperref}
\usepackage{verbatim}
\usepackage{txfonts}
\usepackage{amsmath}
\usepackage{epstopdf}
\usepackage[normalem]{ulem}
\usepackage{verbatim}
\usepackage{xcolor}
\usepackage{bm}

\def\blue{\textcolor{black}}

\def\llh{\textcolor{cyan}}

\begin{document}

\title{Controlling Liouvillian topological phases via Hamiltonian band topology under chiral symmetry}

\author{Shu Long}
\affiliation{Quantum Science Center of Guangdong-Hong Kong-Macao Greater Bay Area (Guangdong), Shenzhen, China}
\affiliation{School of Physics and Astronomy \& Guangdong Provincial Key Laboratory of Quantum Metrology and Sensing, Sun Yat-Sen University (Zhuhai Campus), Zhuhai 519082, China}

\author{Hong-Sen Yin}
\affiliation{School of Physics and Astronomy \& Guangdong Provincial Key Laboratory of Quantum Metrology and Sensing, Sun Yat-Sen University (Zhuhai Campus), Zhuhai 519082, China}

\author{Chao Yang}
\affiliation{Department of Physics, Southern University of Science and Technology, Shenzhen 518055, China}

\author{Sen Mu}\email{senmu@pks.mpg.de}
\affiliation{Max Planck Institute for the Physics of Complex Systems, N\"othnitzer Str. 38, 01187 Dresden, Germany}

\author{Jia-Wei Zhang}\email{zhangjw286@mail.sysu.edu.cn}
\affiliation{School of Physics and Astronomy \& Guangdong Provincial Key Laboratory of Quantum Metrology and Sensing, Sun Yat-Sen University (Zhuhai Campus), Zhuhai 519082, China}

\author{Linhu Li}\email{lilinhu@quantumsc.cn}
\affiliation{Quantum Science Center of Guangdong-Hong Kong-Macao Greater Bay Area (Guangdong), Shenzhen, China}

\begin{abstract}
\blue{We establish a direct connection between Hamiltonian band topology and the Liouvillian spectral winding of open quantum systems with quadratic dissipations. This allows the band topology to act as a knob for controlling Liouvillian topology and corresponding non-equilibrium dynamics. In particular, we show analytically that, for chiral Hamiltonians under sublattice-unidirectional chiral dissipation, the Liouvillian spectral winding depends on the Hamiltonian solely through its band winding number, thereby enabling direct band-topological control of the Liouvillian skin effect. We further identify two distinguished coherent and incoherent Liouvillian skin effects therein: the former hosts a band-topological dark state of the Hamiltonian, whereas the latter yields a mixed steady state driven by dissipative nonreciprocity. Our results establish a systematic framework for topological control of spectral and spatial organization in open quantum systems and provide a unified perspective on topology in Hamiltonian and dissipative dynamics.}
\end{abstract}

\maketitle

\emph{Introduction.}---
Open quantum systems can be described by the Lindblad master equation, whose Liouvillian superoperator is intrinsically non-Hermitian and admits complex spectra absent in closed systems~\cite{10.1063/1.522979,lindblad1976generators,PhysRevResearch.4.023036, PRXQuantum.5.030304, PhysRevB.107.035113, Bardyn_2013,PhysRevLett.127.250402,PhysRevResearch.5.043004,PhysRevB.106.035408}.
In particular, a new class of point-gap topological phases characterized by the spectral winding has recently been identified  for such spectra~\cite{PhysRevX.8.031079,PhysRevLett.124.086801,PhysRevLett.125.126402,koch2025liouvillian}.
This point-gap topology is closely connected to the Liouvillian skin effect (LSE), i.e., the accumulation of a macroscopic set of spatial decay modes near boundaries under open boundary conditions (OBC), potentially including the steady state
~\cite{PhysRevLett.123.170401,PhysRevLett.127.070402, PhysRevResearch.4.023160, PhysRevB.108.155114, PhysRevB.110.045440, 6msp-vbbm, Sannia:25, PhysRevResearch.6.L032067, PhysRevB.108.054313,PhysRevB.109.014313,PhysRevB.111.L060301,hu2025manybody,koch2025liouvillian,chaduteau2026lindbladian,sim2025observables}.

Unlike Hamiltonian band topology formulated with Bloch eigenstates and protected by global discrete symmetries~\cite{hasan2010colloquium,qi2011topological,chiu2016classification}, Liouvillian spectral topology is rooted in the complex spectrum of the full Liouvillian superoperator and depends on both coherent Hamiltonian dynamics and dissipative jump processes. 
The two notions of topology are therefore defined at fundamentally different levels, one in the pseudospin space for the Bloch vector manifold of a Hermitian band structure, and the other in the complex plane for eigenvalues of a non-Hermitian operator. 
\blue{Although Liouvillian topology has been extensively explored within effective non-Hermitian band descriptions~\cite{PhysRevLett.123.170401,PhysRevLett.124.040401,PhysRevResearch.4.023160,chaduteau2026lindbladian} and interacting many-body non-Hermitian frameworks~\cite{vznidarivc2015relaxation,medvedyeva2016exact}, \blue{these studies primarily classify the Liouvillian independently of the band topology carried by the coherent Hamiltonian of the same open system. Whether, and under what conditions, Hamiltonian band topology can directly determine Liouvillian spectral topology has therefore remained an open question.}}

In this letter, we establish a \blue{symmetry-enabled} correspondence between Hamiltonian band topology and Liouvillian spectral topology in a class of one-dimensional dissipative lattices, through which the former serves as a knob for tuning the spectral and spatial organization of open quantum systems.
\blue{We analytically demonstrate that the winding topology of the one-dimensional two-band Hamiltonian can be mapped onto a spectral topology of the Liouvillian spectrum under sublattice-unidirectional chiral dissipation.}
Consequently, the LSE and its associated dynamical signatures can be predicted directly from the Hamiltonian band topology, despite the fundamentally non-unitary nature of the Lindblad evolution.
\blue{Furthermore, we unveil coherent and incoherent Liouvillian skin states associated with nontrivial and trivial Hamiltonian band topology, respectively. We also note that, under chiral-asymmetric dissipation, Hamiltonian band topology and Liouvillian spectral topology become independent.}
Our results establish a \blue{symmetry-enabled} bridge between Hamiltonian and Liouvillian topology, revealing a mechanism for controlling non-equilibrium dynamics of open quantum systems via Hamiltonian band topology.

\emph{Model and topological winding numbers.}---
We consider generic one-dimensional two-band models with chiral symmetry, described by the Hamiltonian
\begin{equation}
    \hat{H}=\sum_{l=1}^L\sum_{s}J_{s} |a,l+s\rangle\langle b,l|+{\rm h.c.},
   \label{eq:H}    
\end{equation}
where $l$ labels the unit cell, $L$ is the system size,
and $|s|<L$ denotes the range of hopping from pseudospin or sublattice $b$ to $a$. 
The explicit symmetry condition will be introduced later.
Within the Born-Markov approximation~\cite{Daley04032014,Sieberer_2016}, dynamics of an open system is described by the Lindblad master equation~\cite{10.1063/1.522979,lindblad1976generators},
  \begin{equation}
    \frac{{\rm d}\hat{\rho}}{{\rm d}t}=-i[\hat{H},\hat{\rho}]+\sum_{p}\left(\hat{D}_{p}\hat{\rho}\hat{D}_{p}^\dagger-\frac{1}{2}\{\hat{D}_{p}^\dagger\hat{D}_{p},\hat{\rho}\}\right)\equiv\mathcal{L}{\rho},
   \label{eq:mastereq}
  \end{equation}
where $\hat{\rho}$ is the density matrix and $\hat{D}_p$ are the jump operators. We take quadratic dissipators of the form
\begin{equation}        
\hat{D}_{p}\rightarrow \hat{D}_{l,s}^{\alpha,\alpha'}=\sqrt{\gamma_{s}^{\alpha,\alpha'}} 
|\alpha, l+s \rangle\langle \alpha',l |
\label{eq:D}    
\end{equation}
with $\alpha, \alpha'\in \{a,b\}$, representing structured single-particle jump processes. 
The Liouvillian $\mathcal{L}$ acts on the vectorized density matrix $\rho$ in an extended Hilbert space of dimension $4L^2$ (see Supplemental Materials~\cite{SuppMat}), and satisfies $\mathcal{L}\rho_j=\lambda_j\rho_j$, where $\lambda_j$ are the eigenvalues and $\rho_j$ the corresponding normalized eigenmodes. The steady state is denoted by $\rho_0$ with $\lambda_0=0$, and the Liouvillian gap is defined as 
$
\Delta\lambda=|{\rm Re}[\lambda_1]|
$
where ${\rm Re}[\lambda_1]\geq {\rm Re}[\lambda_j]$ for all $j\neq 0$.

\blue{Topological invariants are typically defined for translationally invariant Bloch Hamiltonians and Liouvillians under periodic boundary condition (PBC).}
By applying the Fourier transformation 
$   
       |\alpha,l\rangle\langle \alpha',l'|=\frac{1}{L}\sum_{k,k'} e^{-ikl}e^{ik'l'}|\alpha,k\rangle\langle \alpha'k'|
$,
we obtain the Bloch Hamiltonian in Eq.~\eqref{eq:H},
   \begin{equation}\label{eq:hk}
   H(k)=\begin{pmatrix}
    0&{h}^*_k\\
    {h}_k&0
   \end{pmatrix}   
  \end{equation}
with ${h}_k=\sum_{s}^{} J^*_{s}e^{{i}sk}$. 
The  chiral symmetry reads
\begin{align}
\{\sigma_z ,H(k)\}=0,\label{eq:chiral}
\end{align} 
with $\sigma_z$ acting in the pseudospin space. We can then define a winding number $W_H$ characterizing the band topology of $\hat{H}$,
\begin{equation}
W_H=\frac{1}{2\pi i}\int_0^{2\pi} {\rm d} k\frac{{\rm d}}{{\rm d}k}\ln {h}_k.
\label{eq:WH}
\end{equation}
By definition, $W_H$ can take arbitrary integers, representing a $\mathbb{Z}$-type topology. Notably, each hopping term $J_s$ contributes a winding number $s$ when considered alone, while the total winding is a result of the competition between Fourier harmonics in ${h}$.

For the Liouvillian superoperator $\mathcal{L}$ in Eq.~\eqref{eq:mastereq},
although different quasi-momenta $k$ and $k'$ are coupled by the dissipative term $\hat{D}_p\rho\hat{D}_p^\dagger$, the momentum difference $K=k'-k$ labels an invariant subspace of the Liouvillian (see Supplemental Materials~\cite{SuppMat}).
Thus, to describe the Liouvillian topology associated with the steady-state at $\lambda_0=0$, we define a spectral winding number for $\mathcal{L}(K)$ as
\begin{equation}
W_0=\frac{1}{2\pi i}\int_0^{2\pi} {\rm d} K\frac{{\rm d}}{{\rm d}K}\ln\det[\mathcal{L}(K)-0^-],
\label{eq:winding}
 \end{equation}
which \blue{also represents a $\mathbb{Z}$-type topology and} characterizes the winding of the Liouvillian spectrum near $\lambda=0$.
\blue{Note that, due to the opposite signs of the momenta $k$ and $k'$ associated with the ket and bra, the conserved $K=k'-k$ serves as the Bloch momentum of an effective 1D Liouvillian, while the remaining $(k,k')$ modes within each $K$ sector constitute its internal degrees of freedom. Consequently, $W_0$ is precisely the spectral winding number of this effective 1D Bloch Liouvillian.}
Given the established bulk-boundary correspondence between the non-Hermitian skin effect and non-trivial spectral winding~\cite{PhysRevLett.124.056802,PhysRevLett.124.086801,PhysRevLett.125.126402}, 
\blue{one expects localization along the diagonal direction of $\rho_0$, representing the LSE of steady states, whenever $W_0\neq0$.}

\emph{\blue{Symmetry-enabled} topological imprint of the Hamiltonian onto Liouvillian spectral winding.}---
We aim to identify regimes where the topologies of the Hamiltonian and the Liouvillian can be captured by a symmetry-based framework. To achieve this, we consider dissipation that preserves the same chiral symmetry as the Hamiltonian in Eq.~\eqref{eq:chiral}, $\{\sigma_z,\hat{D}_p\}=0$, 
\blue{or equivalently $\alpha'\neq\alpha$ in Eq.~\eqref{eq:D}.
We further consider quantum jump processes that are unidirectional between the two sublattices
and, without loss of generality, take $\alpha'=b$ and $\alpha=a$.
}
\blue{Under such conditions, 
we analytically find that the PBC steady state appears at $K=0$, and the eigenvalue deviated from it satisfies~\cite{SuppMat}
\begin{align}
\Delta\lambda_0\propto iK\sum_s(s-W_H)\gamma_s^{a,b}.
\end{align}
 Since all eigenvalues of $\mathcal{L}(K)$ have their real parts being negative,
we can deduce the point-gap winding as
\begin{align}
 W_0={\rm Sgn}[\sum_s (s-W_H)\gamma_s^{a,b}], \label{eq:W_sH}
\end{align}
and thus establish a correspondence between the Hamiltonian band topology and the Liouvillian spectral topology.
}

\blue{
Before turning to explicit examples, we note that the band-topological control of the Liouvillian spectral winding in Eq.~\eqref{eq:W_sH} takes a particularly transparent form when the chiral-symmetric dissipation is unidirectional between the two sublattices, involving either $\gamma_s^{a,b}$ or $\gamma_s^{b,a}$. When dissipative transfer occurs in both sublattice directions, the relation generalizes to a broader form, in which $W_0$ depends on both the Hamiltonian winding $W_H$ and the microscopic hopping parameters $J_s$ (see Supplemental Materials~\cite{SuppMat} for details).
}

\begin{figure}
      \centering
      \includegraphics[width=1\linewidth]{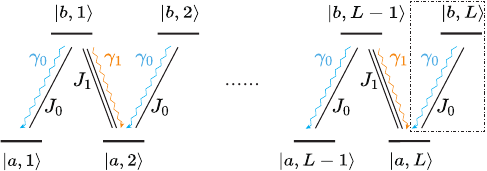}
      \caption{Schematic of dissipative SSH model. $J_0$ (black single lines) and $J_1$ (black double lines) denote coherent hopping amplitudes.
       The blue and orange arrows represent two opposite dissipative couplings, $\gamma_0$ and $\gamma_1$.
       $|\Psi_{\rm left}\rangle$ in Eq.~\eqref{eq:psi_left}
       strictly gives an OBC eigenstate of the Hamiltonian when $J_0<J_1$ and $L\rightarrow \infty$, or in the  edge-defect lattice with $|b,L\rangle$ removed.}
      \label{fig:1}
  \end{figure}

\emph{A dissipative Su-Schrieffer-Heeger model.}---
As a concrete example, we consider a dissipative extension of the Su-Schrieffer-Heeger (SSH) model~\cite{PhysRevLett.42.1698},
given by the Hamiltonian of Eq.~\eqref{eq:H} with nonzero $J_0$ and $J_1$ only \blue{(so that $W_H$ takes only $0$ or $1$)}, and two nearest-neighbor dissipative processes $\gamma_0^{a,b}\equiv\gamma_0$ and $\gamma_1^{a,b}\equiv\gamma_1$, as illustrated in Fig. \ref{fig:1}.
\blue{The Liouvillian winding number is reduced to
\begin{align}
W_0={\rm Sgn}[(1-W_H)\gamma_1-W_H\gamma_0]=\begin{cases}
 1, & \text{if~}W_H=0 \\
 -1, & \text{if~}W_H=1
\end{cases}.
\label{eq:W_SSH}
\end{align}
}
\begin{figure*}
      \centering
      \includegraphics[width=1\linewidth]{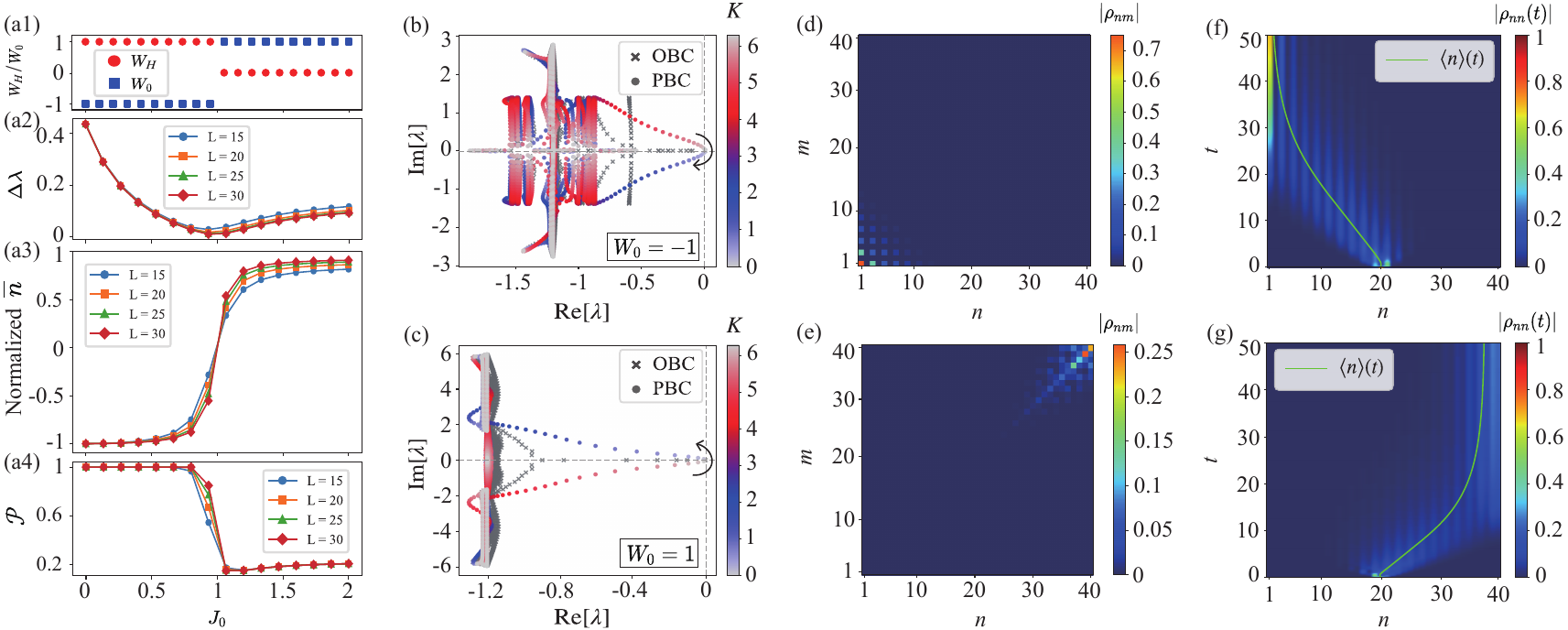}
      \caption{(a) Topological transition of LSE characterized by 
      the band winding number $W_H$ [red dots in (a1)] and Liouvillian spectrum winding number $W_0$ [blue squares in (a1)].
      Under OBC, the transition manifests as the closing of the Liouvillian gap $\Delta\lambda$ (a2),
      the jump of normalized average position $\bar{n}/(L-0.5)$ of the steady state between $-1$ and $1$ (a3), and the jump of steady state purity $\mathcal{P}$ (a4).
      (b) and (c) show the Liouvillian spectrum under OBC (gray) and PBC (colored). 
      Colors represent different $K$ values of the PBC eigenvalues. 
      (d) and (e) the density matrix of the OBC steady-state $\hat{\rho}_{0}$ in (b) and (c), respectively. 
      (f) and (g) time evolution of diagonal elements of the density matrix, $\rho_{nn}$, for an initial state $|\psi_{\rm ini}\rangle=|a,L/2\rangle$.
      Green solid lines indicate the average position $\langle n\rangle=\sum_n n\rho_{nn}$.
      (b), (d), and (f) have $J_0=0.5$, and (c), (e), and (g) have $J_0=2$.
      The other parameters are $J_1=1$, $\gamma_0=\gamma_1=1.2$, and $L=20$.
      }
      \label{fig:2}
  \end{figure*}

In Fig.~\ref{fig:2}(a1), we show numerical results of the Liouvillian and Hamiltonian winding numbers as functions of $J_0$ at fixed $J_1$, both exhibiting a topological transition at $J_0=J_1$. 
\blue{Two illustrative cases of the full Liouvillian spectra are presented in Figs.~\ref{fig:2}(b) and \ref{fig:2}(c). The topological features near $\lambda=0$ are clearly visible: the PBC spectrum forms a loop encircling the OBC spectrum, whose winding direction is consistent with non-trivial point-gap winding number $W_0$.
Under OBC, the system also undergoes a phase transition at this point, evidenced by the asymptotic closing of the Liouvillian gap with increasing system size, as shown in Fig.~\ref{fig:2}(a2). 
This transition can be captured by the steady state distribution $\rho_{nm}$ under OBC, and the average position of its diagonal elements $\bar{n}$,
\begin{align}
\rho_{nm}=\langle n|\hat{\rho}_0|m\rangle,\quad \bar{n}=\sum_n (n-n_0){\rho}_{nn}.
\label{eq:ave_n}
\end{align}
As depicted in Figs.~\ref{fig:2}(d) and \ref{fig:2}(e),  $\rho_{nm}$ shows clear localization at left (right) edge when $W_0=-1$ ($W_0=1$), indicating the LSE corresponding to the Liouvillian spectral winding.
As shown in Fig. \ref{fig:2}(a3), 
the normalized displacement $\bar{n}/(L-0.5)$ switches sharply from $-1$ to $1$ across the transition point $J_1=J_0$, further confirming the correspondence between LSE direction and the sign of $W_0$.}
Beyond static properties, we also probe the dynamics by initializing the system in a localized state at the center, $|\psi_{\rm ini}\rangle=|a,L/2\rangle$. As shown in Figs.~\ref{fig:2}(f) and \ref{fig:2}(g), the time evolution of the diagonal elements $\rho_{nn}(t)$ obtained from Eq.~\eqref{eq:mastereq} shows a unidirectional drift toward the edge aligned with the localization direction of steady states. This dynamical behavior offers an additional clear signature of the LSE, governed by the Hamiltonian band topology.

\blue{
\emph{Band-topological origin of the coherent LSE.}---
Beyond their opposite localization directions, the LSE steady states in Figs.~\ref{fig:2}(d) and \ref{fig:2}(e) exhibit qualitatively distinct coherent and incoherent structures, respectively. We elucidate this difference by considering the topological regime $J_0<J_1$, where the Hamiltonian supports sublattice-polarized band-topological edge states.
In particular, the left-edge state $|\Psi_{\rm left}\rangle\equiv \psi_{a,l}|a,l\rangle+\psi_{b,l}|b,l\rangle$ is given by
\begin{align}
    \psi_{a,l}\propto (-J_0/J_1)^{l-1},~\psi_{b,l}=0\label{eq:psi_left}
\end{align}
in the thermodynamic limit.
With vanishing amplitudes on $b$ sublattice, $|\Psi_{\rm left}\rangle$ 
is unaffected by the dissipation $\gamma_s^{a,b}$, yielding a dark state~\cite{yang2023symmetry,yang2023dissipative,SuppMat}
\begin{align}
\mathcal{L}\rho_{\rm left}=0,\quad\rho_{\rm left}=|\Psi_{\rm left}\rangle\langle \Psi_{\rm left}|.
\end{align}
Thus, the steady state is a coherent pure state given by the topological edge state [as in Fig. \ref{fig:2}(d)].
In contrast, all Hamiltonian eigenstates are unpolarized when $W_H=0$, thus the LSE steady state is mixed and incoherent [as in Fig. \ref{fig:2}(e)].
We quantify this distinction using the purity, defined as
\begin{align}
    \mathcal{P}={\rm Tr}[\hat{\rho}_0^2].
\end{align}
As shown in Fig.~\ref{fig:2}(a4), the purity reveals the qualitatively different organization of the LSE steady state in the regimes with nontrivial and trivial Hamiltonian band topology.
}

\begin{figure}[h]
     \centering
     \includegraphics[width=1\linewidth]{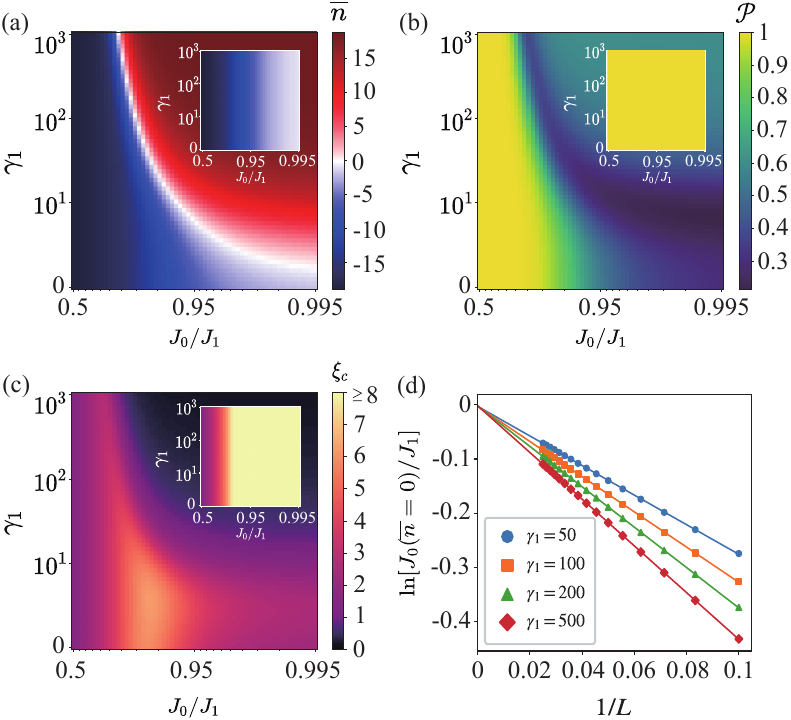}
     \caption{(a) the average position $\overline n$, (b) purity $\mathcal{P}$, and (c) coherence decaying length $\xi_c$ 
     of the steady state under OBC in the same phase with $W_0=-1$
     Insets show the same quantities for the edge-defect lattice with $|b,L\rangle$ removed [the system center in Eq.~\eqref{eq:ave_n} becomes $n_0=L$].
     (d) Scaling of the critical $J_0$, extracted from (a) with $\bar{n}=0$. When $L\rightarrow\infty$, the critical value tends to $J_0=J_1$, the topological transition point under PBC .
     Other parameters are $J_1=2$ and $\gamma_0=2$.
     }
     \label{fig:3}
 \end{figure}

\blue{The preceding analysis is valid in the thermodynamic limit, where the state $|\Psi_{\rm left}\rangle$ becomes an exact Hamiltonian eigenstate. At finite $L$, it hybridizes with the right-edge state polarized on the $b$ sublattice. Consequently, sufficiently strong rightward dissipation, such as that induced by $\gamma_1$, can drive the steady state toward the right boundary even when the bulk invariant $W_0$ predicts a leftward LSE.
}


\blue{
In Fig.~\ref{fig:3}, we characterize the finite-size effect through the spatial distribution, purity, and coherence of the steady state. As shown in Figs.~\ref{fig:3}(a) and \ref{fig:3}(b), increasing $\gamma_1$ drives the steady state toward the right boundary and reduces its purity. 
The same behavior emerges as the system approaches the Hamiltonian topological transition at $J_0=J_1$, where finite-size hybridization between the two edge states becomes increasingly pronounced.
To quantify the coherence of the steady state, we define the coherence-decay length
\begin{align}
\xi_c=1/\ln |C(0)/C(1)|,\quad
C(d)=\sum_{l}  \sum_{\alpha,\alpha'} \langle \alpha, l |\hat{\rho}_0|\alpha',l+d\rangle,
\end{align}
assuming an exponentially decaying distance-resolved coherence profile, $C(d\geq0)\sim e^{-d/\xi_c}$. When $\hat{\rho}_0=\hat{\rho}_{\rm left}$, $\xi_c$ coincides with the localization length of the edge state $|\Psi_{\rm left}\rangle$. By contrast, an incoherent steady state is expected to exhibit a substantially shorter coherence length, as confirmed in Fig.~\ref{fig:3}(c).}


\blue{We note that the agreement between the bulk invariant $W_0$ and the LSE is recovered in two complementary limits. First, the finite-size correction vanishes as $L\rightarrow\infty$, as shown in Fig.~\ref{fig:3}(d). Second, removing the site $(b,L)$ from the open chain ensures that $|\Psi_{\rm left}\rangle$ remains an exact eigenstate even at finite size, thereby stabilizing the dark steady state~\cite{SuppMat}. As demonstrated in the insets of Figs.~\ref{fig:3}(a)-\ref{fig:3}(c), the resulting steady-state distribution, purity, and coherence become insensitive to $\gamma_1$, yielding an exact bulk-boundary correspondence governed by $W_0$.}

\blue{A remark is in order here. We emphasize that Eq.~\eqref{eq:W_sH} establishes the Hamiltonian-Liouvillian topological correspondence under PBC, while the OBC steady-state structure is governed by two distinct boundary-localization mechanisms: dissipative nonreciprocity and the Hamiltonian topological dark state. Their cooperation produces the coherent LSE in Fig.~\ref{fig:2}(d), whereas the incoherent LSE in Fig.~\ref{fig:2}(e) arises solely from dissipative nonreciprocity. More generally, these mechanisms can also compete, giving rise to steady and metastable modes localized at opposite boundaries without altering the PBC topological correspondence, as demonstrated in the Supplemental Material~\cite{SuppMat}.}



\emph{Chiral-asymmetric dissipation.}---
\blue{
To further underscore the essential role of chiral symmetry in enabling the band-topology control of Liouvillian spectral winding and the LSE, we next consider other dissipation terms $\gamma_{1}^{a,a}$ and $\gamma_{-1}^{b,b}$, for which $\{\sigma_z,\hat{D}_p\}\neq 0$. With this chiral-asymmetric dissipation, the Liouvillian winding number becomes~\cite{SuppMat},
\begin{align}
W_0={\rm Sgn}[\sum_s s\gamma_s^{a,a}
+ \sum_s s\gamma_s^{b,b}],
\end{align}
and is determined entirely by the dissipative channels, independently of the Hamiltonian band topology.}

\begin{figure}
      \centering
      \includegraphics[width=1\linewidth]{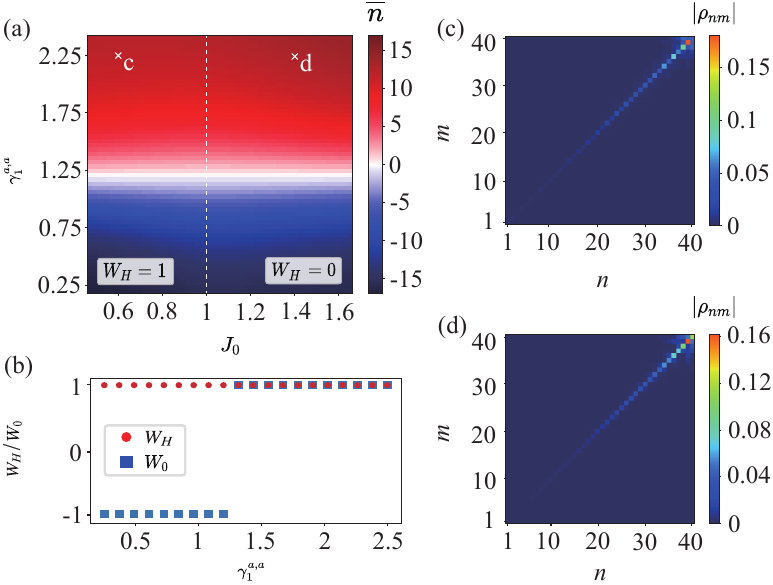}
      \caption{ Phase diagram and LSE with chiral-asymmetric dissipation. (a) The average position $\overline{n}$ for different values of $J_0$ and $\gamma_1^{a,a}$.
      White dashed line marks the transition of the band topology characterized by $W_H$. (b) Hamiltonian winding number $W_H$ (red dots) and Liouvillian winding number $W_0$ (blue squares) for different values of $\gamma_1^{a,a}$. (c) and (d) the density matrix of the OBC steady-state $\hat{\rho}_{0}$ with $J_0=0.6$ and $J_0=1.4$, respectively, marked by the white crosses in (a). The same localization direction is observed despite that they possess different $W_H$. 
      The other parameters are $J_1=1$, $\gamma_{-1}^{b,b}=1.2$,  $L=20$; and $\gamma_1^{a,a}=2.25$ in (b) and (c).}
      \label{fig:4}
\end{figure}

In Fig.~\ref{fig:4}(a), we present the resulting phase diagram under chiral-asymmetric dissipation. In stark contrast to the symmetric case, the band topology no longer controls the direction of the LSE; instead, the LSE direction is dictated solely by the relative strengths of the dissipative processes. Consistently, the Liouvillian winding number $W_0$ also becomes independent of the Hamiltonian winding number $W_H$ [Fig. \ref{fig:4}(b)].
The steady-state density matrices $\hat{\rho}_{0}$ of two representative cases with different $W_H$ are shown in Figs. ~\ref{fig:4}(c) and (d), and both exhibit localization toward the same boundary, confirming the loss of topological correspondence. 
This behavior can be understood by noting that, under chiral-symmetric dissipation, the band topology effectively acts as a gauge field that modifies $W_0$ through a unitary transformation, thereby reversing the LSE direction. 
However, for the present asymmetric dissipation $\gamma_{1}^{a,a}$ and  $\gamma_{-1}^{b,b}$, the winding information resides in diagonal components that are insensitive to this gauge structure (see SM~\cite{SuppMat}). 
Consequently, the Hamiltonian band topology no longer influences the direction of the LSE, underscoring that symmetry alignment is essential for the topological imprinting mechanism established in the main text.

\emph{Summary and discussion.}---
\blue{We have established a correspondence between the Hamiltonian band topology and the Liouvillian spectral topology, as well as its further connection to the LSE: the shared chiral symmetry of the Hamiltonian and dissipation gives a prerequisite for the two levels of topology to be intertwined, while the unidirectionality of dissipation ensures an exact topological correspondence between them.}
\blue{The correspondence is analytically obtained and demonstrated for two-band single-particle systems; 
and we further provide numerical evidence that the correspondence persists in representative multiband and multiparticle systems in Supplemental Materials~\cite{SuppMat}.
Furthermore, we identify different organizations of the LSE steady state, either as a coherent pure state from the Hamiltonian band-topology, or an incoherent mixed one.}
These findings not only bridge the two distinct topological structures through symmetry protection, but also unveil the closed Hamiltonian topology as a governing principle and controlling method for the spectral and spatial organization of open quantum dynamics.

\blue{From a microscopic point of view, the sublattice-unidirectional dissipation necessary for the exact topological correspondence can arise from an amplitude-damping mechanism, where the $b$-sublattice state is coupled to short-lived excited states and subsequently decays into the $a$-sublattice states.}
Experimentally, the Hamiltonian and unidirectional dissipation in our model may be implemented with ultracold atoms in optical lattices~\cite{ Jaksch_2003, PhysRevLett.111.185301, PhysRevLett.111.185302, JDalibard_1985, PhysRevA.82.063605}, trapped-ion setups~\cite{PhysRevLett.77.4728,harrington2022engineered} and superconducting qubits~\cite{mi2024stable}, offering a route for realizing topologically programmed steady states through our formalism.
\blue{For example, trapped-ion platforms are an ideal candidate for realizing this chiral-symmetric dissipation. By coupling internal and external states to form a phonon chain, one can construct two sublattices as $\lvert e,n\rangle$ and $\lvert g,n\rangle$ (with $n$ the phonon number). The coherent couplings $J_0$ and $J_1$ are implemented via carrier and red-sideband transitions, respectively, while the dissipative terms $\gamma_0$ and $\gamma_1$ are engineered by coupling the relevant levels to short-lived excited states and employing adiabatic elimination: $\gamma_0$ can be realized via a carrier coupling \cite{zhang2020single,zhang2022dynamical,wu2026experimental} and $\gamma_1$ via a red-sideband coupling.}

\emph{Acknowledgment.}--- 
 We acknowledge helpful discussion with Yu-Min Hu.
This work is supported by the National Natural Science Foundation of China (Grants No. 12474159 and No. 12304315) and the Guangdong Provincial Quantum Science Strategic Initiative (Grants No. GDZX2504003 and GDZX2504006).

\clearpage
\onecolumngrid

\begin{center}
    \textbf{\large Supplemental Materials}
\end{center}

\tableofcontents
\setcounter{secnumdepth}{2}
\setcounter{equation}{0} \setcounter{figure}{0} 
\setcounter{table}{0} %
\renewcommand{\theequation}{S\arabic{equation}} 
\renewcommand{\thefigure}{S%
\arabic{figure}}

\section{Matrix form of the Liouvillian superoperator}

  In this section, we derive in detail the explicit expression for the corresponding Liouvillian superoperator after vectorizing the density matrix as
  \begin{equation}   
    |\alpha,l\rangle\langle \alpha',l'|\rightarrow |\alpha,\alpha',l,l'\rangle\rangle.
   \label{eq:vectorS}     
  \end{equation}
For convenience, we derive the three terms in the Lindblad master equation [Eq. (2) of the main text],
  \begin{equation}
    \frac{{\rm d}\hat{\rho}}{{\rm d}t}=-i[\hat{H},\hat{\rho}]+\sum_{p}\left(\hat{D}_{p}\hat{\rho}\hat{D}_{p}^\dagger-\frac{1}{2}\{\hat{D}_{p}^\dagger\hat{D}_{p},\hat{\rho}\}\right)\equiv\mathcal{L}{\rho},
   \label{eq:mastereq}
  \end{equation}
separately.
  
 \subsection{Hamiltonian term}
 
  We first consider two-band systems with nontrivial topological properties protected by a chiral symmetry. The Hamiltonian is given by
  \begin{equation}
    \hat{H}=\sum_{l=1}^L\sum_{s} J_{s} |a,l+s\rangle\langle b,l|+{\rm h.c.},
   \label{eq:HS}    
  \end{equation}
  The Hamiltonian term in the Liouvillian superoperator can be expressed as
  \begin{equation}
   \begin{aligned}
    \mathcal{L}_{H}(\hat{\rho})=-i[\hat{H},\hat{\rho}]&=-i\sum_{l,l'}\bigg(\big[\hat{H},|b,l\rangle \langle b,l'|\big]+\big[\hat{H},|a,l\rangle \langle a,l'|\big]+\big[\hat{H},|b,l\rangle \langle a,l'|\big]+\big[\hat{H},|a,l\rangle \langle b,l'|\big]\bigg)\\
    &=A+B+C+D,   
   \end{aligned}
  \end{equation}
  with
  \begin{equation}
    A=-i\sum_{l,l'}\bigg(\sum_{s}^{}J_{s}|a,l+s\rangle\langle b,l'|
     -\sum_{s}^{} J_{s}^*|b,l\rangle\langle a,l'+s|\bigg),
  \end{equation}
  \begin{equation}
    B=-i\sum_{l,l'}\bigg(\sum_{s}^{}J_{s}^*|b,l-s\rangle\langle a,l'|
     -\sum_{s}^{}J_{s}|a,l\rangle\langle b,l'-s|\bigg),
  \end{equation}
  \begin{equation}
    C=-i\sum_{l,l'}\bigg(\sum_{s}^{}J_{s}|a,l+s\rangle\langle a,l'|
     -\sum_{s}^{}J_{s}|b,l\rangle\langle b,l'-s|\bigg),
  \end{equation}
  \begin{equation}
    D=-i\sum_{l,l'}\bigg(\sum_{s}^{} J_{s}^*|b,l-s\rangle\langle b,l'| 
     -\sum_{s}^{}J_{s}^*|a,l\rangle\langle a,l'+s|\bigg).
  \end{equation}
  In the vector basis, these terms read:
  \begin{equation}
    \mathcal{L}_{H}|b,b,l,l'\rangle\rangle=-i\bigg(\sum_{s}^{}J_{s}|a,b,l+s,l'\rangle\rangle
     -\sum_{s}^{}J_{s}^*|b,a,l,l'+s\rangle\rangle\bigg),
  \end{equation}
  \begin{equation}
    \mathcal{L}_{H}|a,a,l,l'\rangle\rangle=-i\bigg(\sum_{s}^{}J_{s}^*|b,a,l-s,l'\rangle\rangle 
     -\sum_{s}^{}J_{s}|a,b,l,l'-s\rangle\rangle\bigg),
  \end{equation}
  \begin{equation}
    \mathcal{L}_{H}|b,a,l,l'\rangle\rangle=-i\bigg(\sum_{s}^{}J_{s}|a,a,l+s,l'\rangle\rangle
     -\sum_{s}^{}J_{s}|b,b,l,l'-s\rangle\rangle\bigg),
  \end{equation}
  \begin{equation}
    \mathcal{L}_{H}|a,b,l,l'\rangle\rangle=-i\bigg(\sum_{s}^{} J_{s}^*|b,b,l-s,l'\rangle\rangle  
     -\sum_{s}^{}J_{s}^*|a,a,l,l'+s\rangle\rangle\bigg).
  \end{equation}
  Thus $\mathcal{L}_H$ can be written as
  \begin{equation}
   \begin{aligned}
    \mathcal{L}_H=-i\sum_{l,l'}\Bigg[
     \sum_{s}^{}J_{s}\bigg(
     &|a,b,l+s,l'\rangle\rangle\langle\langle b,b,l,l'|
     +|a,a,l+s,l'\rangle\rangle\langle\langle b,a,l,l'|\\
     &-|a,b,l,l'-s\rangle\rangle\langle\langle a,a,l,l'|
     -|b,b,l,l'-s\rangle\rangle\langle\langle b,a,l,l'|
       \bigg)+{\rm h.c.}\Bigg].
   \end{aligned}
   \label{eq:matrixLH}
  \end{equation}

 \subsection{First term of jump operators}
  Consider the jump operators given by:
  \begin{equation}
    \hat{D}_{l,s}^{\alpha,\alpha'}=\sqrt{\gamma_{s}^{\alpha,\alpha'}} |\alpha, l+s \rangle\langle \alpha',l |.
   \label{eq:jumpopsS}
  \end{equation}
  The first quantum jump term in Eq.~(\ref{eq:mastereq}) is given by
  \begin{equation}
   \begin{aligned}
    \mathcal{L}_{D1}(\hat{\rho})&=\sum_{l}\sum_{s,\alpha,\alpha'}
     \hat{D}_{l,s}^{\alpha,\alpha'}\rho [\hat{D}_{l,s}^{\alpha,\alpha'}]^\dagger\\
     &=\sum_{l}\sum_{s,\alpha,\alpha'} \gamma_s^{\alpha,\alpha'}|\alpha, l+s \rangle\langle \alpha',l | \hat{\rho} |\alpha', l \rangle\langle \alpha,l+s |,   
   \end{aligned}    
  \end{equation}
  which leads to
  \begin{equation}
   \mathcal{L}_{D1}|\alpha',\alpha',l,l\rangle\rangle=\sum_{s,\alpha}
   \gamma_s^{\alpha,\alpha'}|\alpha,\alpha,l+s,l+s\rangle\rangle.
  \end{equation}
  The operator can be written as
  \begin{equation}
    \mathcal{L}_{D1}=\sum_{l}\sum_{s,\alpha,\alpha'}\gamma_s^{\alpha,
    \alpha'}|\alpha,\alpha,l+s,l+s\rangle\rangle\langle\langle \alpha',\alpha',l,l|.  
   \label{eq:matrixLD1}
  \end{equation}

 \subsection{Second term of jump operators}

  According to Eq.~(\ref{eq:jumpopsS}), the second term of the jump operators in Eq.~(\ref{eq:mastereq}) is:
  \begin{equation}
   \begin{aligned}
    \mathcal{L}_{D2}(\hat{\rho})&=-\frac{1}{2}\sum_{l}\sum_{s,\alpha,\alpha'}\{[\hat{D}_{l,s}^{\alpha,\alpha'}]^\dagger \hat{D}_{l,s}^{\alpha,\alpha'},\hat{\rho}\}\\
    &=-\frac{1}{2}\sum_l\sum_{s,\alpha,\alpha'}\gamma_{s}^{\alpha,\alpha'}\{|\alpha',l\rangle\langle \alpha',l|,\hat{\rho}\}.   
   \end{aligned}    
  \end{equation}
  Acting on different terms of the density matrix, we obtain
  \begin{equation}
    \mathcal{L}_{D2}|\alpha',\alpha',l,l'\rangle\rangle=-\sum_{s,\alpha}\gamma_{s}^{\alpha,\alpha'}|\alpha',\alpha',l,l'\rangle\rangle,
  \end{equation}
  \begin{equation}
    \mathcal{L}_{D2}|b,a,l,l'\rangle\rangle=-\frac{1}{2}\sum_{s,\alpha}\gamma_s^{\alpha,b}|b,a,l,l'\rangle\rangle
    -\frac{1}{2}\sum_{s,\alpha}\gamma_s^{\alpha,a}|b,a,l,l'\rangle\rangle,
  \end{equation}
  \begin{equation}
    \mathcal{L}_{D2}|a,b,l,l'\rangle\rangle=-\frac{1}{2}\sum_{s,\alpha}\gamma_s^{\alpha,a}|a,b,l,l'\rangle\rangle
    -\frac{1}{2}\sum_{s,\alpha}\gamma_s^{\alpha,b}|a,b,l,l'\rangle\rangle.    
  \end{equation}
  Thus the matrix form is given by
  \begin{equation}
   \begin{aligned}
    \mathcal{L}_{D2}=-\sum_{l,l'}\bigg[&\sum_{s,\alpha}\bigg(\gamma_s^{\alpha,b}|b,b,l,l'\rangle\rangle\langle\langle b,b,l,l'|+\gamma_s^{\alpha,a}|a,a,l,l'\rangle\rangle\langle\langle a,a,l,l'|\bigg)\\\
    &+\frac{1}{2}\sum_{s,\alpha}\bigg(\gamma_s^{\alpha,b}+\gamma_s^{\alpha,a}\bigg)\bigg(|b,a,l,l'\rangle\rangle\langle\langle b,a,l,l'|+|a,b,l,l'\rangle\rangle\langle\langle a,b,l,l'|\bigg)
     \bigg],   
   \end{aligned} 
   \label{eq:matrixLD2}
  \end{equation}

  Combining Eq.~(\ref{eq:matrixLH}), Eq.~(\ref{eq:matrixLD1}), and Eq.~(\ref{eq:matrixLD2}) yields the matrix form of the Liouvillian superoperator in real space. Diagonalizing this matrix gives the Liouvillian spectrum and its eigenmodes.  

\section{The Liouvillian superoperator in momentum space}

  After the Fourier transformation, Eq.~(\ref{eq:matrixLH}) can be written as
  \begin{equation}
   \begin{aligned}
    \mathcal{L}_H(k,k')=-i\sum_{s} \bigg[ J_{s}\bigg(&e^{-isk}|a,b,k,k'\rangle\rangle\langle\langle b,b,k,k'|-e^{-isk'}|b,b,k,k'\rangle\rangle\langle\langle b,a,k,k'|\\
    &+e^{-isk}|a,a,k,k'\rangle\rangle\langle\langle b,a,k,k'|-e^{-isk'}|a,b,k,k'\rangle\rangle\langle\langle a,a,k,k'|\bigg)+{\rm h.c.}\bigg]. 
   \end{aligned} 
   \label{eq:L_H_k}
  \end{equation}
  Its matrix form, with the basis arranged in the order of $(aa,ab,ba,bb)$, reads
  \begin{equation}
    \begin{pmatrix}
     0& ih_{k'} & -ih^*_k & 0\\
     ih^*_{k'} & 0& 0 & -ih^*_k\\
     - ih_k & 0 &0 & ih_{k'}\\
     0 & -ih_k & ih^*_{k'}& 0
    \end{pmatrix},
   \label{eq:L_H_k_matrix}
  \end{equation}
  with $h_k=\sum_{s}^{}J^*_{s}e^{isk}$.

  Eq.~(\ref{eq:matrixLD1})'s Bloch form after the transformation $|\alpha,\alpha',l,l'\rangle\rangle=\frac{1}{L}\sum_{k,k'} e^{-ikl}e^{ik'l'}|\alpha,\alpha',k,k'\rangle\rangle$ becomes more complicated:
  \begin{equation}
   \begin{aligned}
    \mathcal{L}_{D1}(k,k')&=\sum_{s,\alpha,\alpha'}\frac{\gamma_{s}^{\alpha,\alpha'}}{L^2}\sum_{k_1,k_1'}\sum_{k_2,k_2'}\sum_l e^{-i(k_1-k_1'-k_2+k_2')l}e^{-i(k_1-k_1')s}|\alpha,\alpha,k_1,k_1'\rangle\rangle\langle\langle \alpha',\alpha',k_2,k_2'|\\
    &=\sum_{s,\alpha,\alpha'}\frac{\gamma_{s}^{\alpha,\alpha'}}{L}\sum_{k_1,k_1'}\sum_{k_2,k_2'}e^{-i(k_1-k_1')s}|\alpha,\alpha,k_1,k_1'\rangle\rangle\langle\langle \alpha',\alpha',k_2,k_2'|\delta_{k_1-k_1'-k_2+k_2'}. 
   \end{aligned} 
   \label{eq:L_D1_k}
  \end{equation}
  With these terms, different "momenta" are coupled together and no longer good quantum numbers, and we have to work in the subspace with constant $K=k'-k$.  Its matrix form, with the basis arranged as ($aa,ab,ba,bb$), reads
  \begin{equation}
    \begin{pmatrix}
    A_{11} &0 & 0 & A_{14}\\
    0 & 0& 0 & 0\\
    0 & 0 & 0 & 0\\
    A_{41} & 0& 0& A_{44}
    \end{pmatrix},  
   \label{eq:L_D1_k_matrix}
  \end{equation}
  where
   $A_{11}=\sum_s\gamma_s^{a,a} e^{iKs}/L$,
   $A_{14}=\sum_s\gamma_s^{a,b} e^{iKs}/L$,
   $A_{41}=\sum_s\gamma_s^{b,a} e^{iKs}/L$,
   $A_{44}=\sum_s\gamma_s^{b,b} e^{iKs}/L$.

  Applying the same transformation to Eq.~(\ref{eq:matrixLD2}) yields:
  \begin{equation}
   \begin{aligned}
    \mathcal{L}_{D2}(k,k')=&-\bigg[\sum_{s,\alpha}\gamma_s^{\alpha,b}|b,b,k,k'\rangle\rangle\langle\langle b,b,k,k'|+\sum_{s,\alpha}\gamma_s^{\alpha,a}|a,a,k,k'\rangle\rangle\langle\langle a,a,k,k'|\\
    &+\frac{1}{2}\sum_{s,\alpha}\bigg(\gamma_s^{\alpha,b}+\gamma_s^{\alpha,a}\bigg)\bigg(|b,a,k,k'\rangle\rangle\langle\langle b,a,k,k'|+|a,b,k,k'\rangle\rangle\langle\langle a,b,k,k'|\bigg)
    \bigg].   
   \end{aligned}
   \label{eq:L_D2_k}
  \end{equation}
  Thus the matrix form is given by
  \begin{equation}
   \begin{pmatrix}
     -\sum_{s,\alpha}\gamma_s^{\alpha,a} &0 & 0 & 0\\
     0 & -\frac{1} {2}\sum_{s,\alpha}\bigg(\gamma_s^{\alpha,b}+\gamma_s^{\alpha,a}\bigg)& 0 & 0\\
     0 & 0 & -\frac{1}  {2}\sum_{s,\alpha}\bigg(\gamma_s^{\alpha,b}+\gamma_s^{\alpha,a}\bigg) & 0\\
     0 & 0& 0& -\sum_{s,\alpha}\gamma_s^{\alpha,b}
    \end{pmatrix}.
   \label{eq:L_D2_k_matrix}
  \end{equation}

  Based on Eq.~(\ref{eq:L_H_k_matrix}), Eq.~(\ref{eq:L_D1_k_matrix}), and Eq.~(\ref{eq:L_D2_k_matrix}), and since $K=k'-k$ is good quantum number, the Liouvillian superoperator in $K$-space takes the form
  \begin{equation}
    \mathcal{L}(K)={\rm diag}[M_1,M_2,...M_L]+N_L\otimes M_0,
    \label{eq:LKS}
  \end{equation}
  with $N_L$ an $L\times L$ matrix with all elements being $1$, and 
  \begin{equation}
    M_n=\begin{pmatrix}
     -B_{1} & ih_{n'} & -ih_n^* & 0\\
     ih^*_{n'} & -(B_{1}+B_{4})/2 & 0 & -ih^*_n\\
     - ih_n & 0 &-(B_{1}+B_{4})/2  & ih_{n'}\\
     0 & -ih_n & ih^*_{n'}& -B_{4}
     \end{pmatrix},
    ~~M_0=\begin{pmatrix}
     A_{11} &0 & 0 & A_{14}\\
     0 & 0& 0 & 0\\
     0 & 0 & 0 & 0\\
     A_{41} & 0& 0& A_{44}
    \end{pmatrix}.
   \label{eq:MNM0_general}
  \end{equation}
  The coefficients in Eq.~(\ref{eq:MNM0_general}) are
  \begin{equation}
   \begin{aligned}
    &A_{11}=\sum_s\gamma_s^{a,a} e^{iKs}/L,~~
    A_{14}=\sum_s\gamma_s^{a,b} e^{iKs}/L,~~
    A_{41}=\sum_s\gamma_s^{b,a} e^{iKs}/L,~~
    A_{44}=\sum_s\gamma_s^{b,b} e^{iKs}/L,\\
    &B_{1}=\sum_{s,\alpha}\gamma_s^{\alpha,a},~~
    B_{4}=\sum_{s,\alpha}\gamma_s^{\alpha,b},~~
    h_n=h(k_n)=\sum_s J^*_s e^{is k_n},
   \end{aligned}
  \end{equation}
  with $k_n=2n\pi/L$ and  $k_n'=K+k_n=K+2n\pi/L$. Note that the continuous momentum $k$ is replaced by discrete one $k_n$ for finite-size systems.

\section{Exact solution of the Liouvillian spectral winding under chiral-symmetric quantum jumps}
We consider jump operators satisfying the same chiral symmetry as the Hamiltonian, i.e.,
\begin{align}
\sigma_z \hat{D}_p\sigma_z=-\hat{D}_p.\label{supp_eq:chiral}
\end{align}
This symmetry condition allows quantum jumps between different sublattices.

\subsection{Unidirectional dissipation ($b\rightarrow a$)}
Here we first consider the case with those from sublattice $b$ to $a$ described by nonzero $\gamma_s^{a,b}$.
The Liouvillian in this case is given by
  \begin{equation}
    \mathcal{L}(K) ={\rm diag}[M_1,M_2,...M_L] + N_L\otimes M_0, 
  \end{equation}
  where
  \begin{equation}
    M_n=\begin{pmatrix}
     -0 & ih_{n'} & -ih_n^* & 0\\
     ih^*_{n'} & -B_{4}/2 & 0 & ih^*_n\\
     - ih_n & 0 &-B_{4}/2  & ih_{n'}\\
     0 & -ih_n & ih^*_{n'}& -B_{4}
     \end{pmatrix},
    \qquad
    M_0=\begin{pmatrix}
    0&0&0&A_{14}\\
    0&0&0&0\\
    0&0&0&0\\
    0&0&0&0
     \end{pmatrix},  
  \end{equation}
with $A_{14}=\sum_s\gamma_s^{a,b} e^{iKs}/L$, $B_4=\sum_{s}\gamma_s^{a,b}$. 
The other case with nonzero $\gamma_s^{b,a}$ can also be solved similarly.
Having both of them will break the topological correspondence between the Liouvillian winding number $W_0$ and the Hamiltonian winding number $W_H$, making the former determined also by the explicit values of hopping parameters, as will be discussed in the next subsection.

\subsubsection{Steady state at $K=0$}
Consider an auxiliary matrices
\begin{equation}
    M_{n,{\rm aux}}=M_n+LM_0=
\begin{pmatrix}
     0 & ih_n & -ih^*_n & B_4\\
     ih^*_n & -B_4/2 & 0 & -ih^*_n\\
     -ih_n & 0 & -B_4/2 & ih_n\\
     0 &  -ih_n & ih^*_n& -B_4
     \end{pmatrix},
\qquad\mathcal{L}_{\rm aux}={\rm diag}[M_{1,{\rm aux}},M_{2,{\rm aux}},...,M_{L,{\rm aux}}],
  \end{equation}
with 
$h_n=h(k_n)=\sum_s J^*_s e^{is k_n}$.
$M_{n,{\rm aux}}$ contains a simple root $\lambda_{\rm aux}=0$,  
\begin{align}
M_{n,{\rm aux}}|\psi_n\rangle\rangle=0,\qquad|\psi_n\rangle\rangle=
\begin{pmatrix}
\psi_{n,1}\\
\psi_{n,2}\\
\psi_{n,3}\\
\psi_{n,4}
\end{pmatrix}=
\begin{pmatrix}
{B_4^2}/{4|h_n|^2}+1\\
{iB_4h^*_n}/{2|h_n|^2}\\
{-iB_4h_n}/{2|h_n|^2}\\
1
     \end{pmatrix}
\qquad\Rightarrow\qquad
\mathcal{L}_{\rm aux}|\Psi\rangle\rangle=0,
\qquad
|\Psi\rangle\rangle=\frac{1}{\sqrt{L}}\begin{pmatrix}
|\psi_1\rangle\rangle\\
|\psi_2\rangle\rangle\\
...\\
|\psi_L\rangle\rangle
\end{pmatrix}.\label{eq:right_ss}
\end{align}
Note that $|\Psi\rangle\rangle$ is not normalized; the factor $1/\sqrt{L}$ is only to remove the size-dependent of its total wave amplitude.
Next, we act the original Liouvillian at $K=0$ on the column vector $|\Psi\rangle\rangle$,
\begin{align}
\mathcal{L}(0)|\Psi\rangle\rangle=|\Phi\rangle\rangle=\frac{1}{\sqrt{L}}\begin{pmatrix}
|\phi_1\rangle\rangle\\
|\phi_2\rangle\rangle\\
...\\
|\phi_L\rangle\rangle
\end{pmatrix},\qquad|\phi_n\rangle\rangle=\begin{pmatrix}
\phi_{n,1}\\
\phi_{n,2}\\
\phi_{n,3}\\
\phi_{n,4}
\end{pmatrix}.
\end{align}
Since $\mathcal{L}(0)$ and $\mathcal{L}_{\rm aux}$  are different only in their $(4n-3,4m)$ elements with $n,m\in\{1,2,...,L\}$, we immediately have
\begin{align}
|\Phi\rangle\rangle&=\mathcal{L}(0)|\Psi\rangle\rangle=\big[\mathcal{L}(0)-\mathcal{L}_{\rm aux}\big]|\Psi\rangle\rangle
=\frac{1}{\sqrt{L}}[(A_{14}L-B_4),0,0,0,(A_{14}L-B_4),0,0,0,...]^T=0,
\end{align}
since $A_{14}L=B_4$ when $K=0$.
Thus we obtain $\mathcal{L}(K=0)||\Psi\rangle\rangle=0$, meaning that $|\Psi\rangle\rangle$ is the steady state with $\lambda_0=0$.

Note that $|\Psi\rangle\rangle$ is a right-eigenvector of the non-Hermitian superoperator $\mathcal{L}$. For later convenience, we also calculate its corresponding left-eigenvector $\langle\langle \Psi'|$, with
\begin{align}
\mathcal{L}^\dagger|\Psi'\rangle\rangle=0.
\end{align}
Similarly, we obtain
\begin{align}
M^\dagger_{n,{\rm aux}}=\begin{pmatrix}
     0 & -ih_n & ih^*_n & 0\\
    -ih^*_n & -B_4/2 & 0 & ih^*_n\\
     ih_n & 0 & -B_4/2 & -ih_n\\
     B_4 &  ih_n & -ih^*_n& -B_4
     \end{pmatrix},\quad
M^\dagger_{n,{\rm aux}}|\psi'_n\rangle\rangle=0,\quad|\psi'_n\rangle\rangle=
\begin{pmatrix}
\psi'_{n,1}\\
\psi'_{n,2}\\
\psi'_{n,3}\\
\psi'_{n,4}
\end{pmatrix}=
\begin{pmatrix}
1\\
0\\
0\\
1
     \end{pmatrix}
~~\Rightarrow~~
\mathcal{L}_{\rm aux}^\dagger|\Psi'\rangle\rangle=0,
\quad
|\Psi'\rangle\rangle=\frac{1}{\sqrt{L}}\begin{pmatrix}
|\psi'_1\rangle\rangle\\
|\psi'_2\rangle\rangle\\
...\\
|\psi'_L\rangle\rangle
\end{pmatrix};\label{eq:left_ss}
\end{align}
\begin{align}
|\Phi'\rangle\rangle&=\mathcal{L}^\dagger(0)|\Psi'\rangle\rangle=\big[\mathcal{L}^\dagger(0)-\mathcal{L}^\dagger_{\rm aux}\big]|\Psi'\rangle\rangle
=\frac{1}{\sqrt{L}}[0,0,0,(A^*_{14}L-B_4),0,0,0,(A^*_{14}L-B_4),...]^T=0,
\end{align}
since $A^*_{14}L-B_4=0$ when $K=0$.
The left-eigenvector of the steady state is thus given by $|\Psi'\rangle\rangle$.

\subsubsection{Deviating from $K=0$}
For a small but nonzero $K=k'-k$,
The Liouvillian can be written as
\begin{align}
\mathcal{L}(K)=\mathcal{L}(0)+\Delta\mathcal{L},\qquad
\Delta \mathcal{L}={\rm diag}[\Delta M_1,\Delta M_2,...,\Delta M_L]+N_L\otimes \Delta M_0,
\end{align}

 \begin{equation}
    \Delta M_n(K)=\begin{pmatrix}
     0 & -K\sum_s sJ^*_s e^{isk_n}& 0 & 0\\
     K\sum_s sJ_s e^{-isk_n}& 0 & 0 & 0\\
     0 & 0 & 0 & -K\sum_s sJ^*_s e^{isk_n}\\
     0 & 0 & K\sum_s sJ_s e^{-isk_n} & 0
     \end{pmatrix},
    \qquad
    \Delta M_0(K)=\begin{pmatrix}
    0&0&0&iK\sum_s s\gamma_s^{a,b}/L\\
    0&0&0&0\\
    0&0&0&0\\
    0&0&0&0
     \end{pmatrix}.
  \end{equation}
The first-order perturbation to the steady-state eigenvalue $\lambda_0=0$ satisfies (note that the left- and right-eigenvectors are not normalized)
\begin{align}
&\Delta \lambda_0 \propto \langle\langle \Psi'| \Delta\mathcal{L}|\Psi\rangle\rangle
=\frac{K}{L}\sum_n \Bigg(
\frac{-iB_4h^*_n}{2|h_n|^2}\sum_s sJ^*_se^{is k_n} +i\sum_s s\gamma_s^{a,b}
-\frac{iB_4h_n}{2|h_n|^2}\sum_s sJ_se^{-isk_n}\Bigg)\\
&=i\frac{K}{L}\sum_n\Bigg(\sum_s s\gamma_s^{a,b}
+ {\rm Im} C_n\Bigg) ,\label{eq:perturbed}
\end{align} 
with $$C_n=\frac{-iB_4h^*_n}{|h_n|^2}\sum_s sJ^*_se^{is k_n}.$$
We can see that $\Delta \lambda_0$ is purely imaginary. Finally, note that the summation runs from $n=1$ to $n=L$, corresponding to $k_n$ varying from $0$ to $2\pi$. 
Furthermore, note that
\begin{align}
h_n=\sum_s J^*_s e^{is k_n}\Rightarrow \frac{d h_n}{d k_n}=i\sum_s J_s^* s e^{is k_n} \Rightarrow C_n=-B_4 \frac{h^*_n (d h_n/dk_n)}{|h_n|^2}.
\end{align}
Thus we have (with $h_n\rightarrow h_k=\sum_s J^*_s e^{is k}$)
\begin{align}
\Delta \lambda_0&\propto i\frac{K}{L}\sum_n\Bigg(\sum_s s\gamma_s^{a,b}
+ {\rm Im} C_n\Bigg)=iK\Bigg[\sum_s s\gamma_s^{a,b}-{\rm Im}\bigg(\sum_{n=1}^{L} B_4 \frac{h^*_n (d h_n/dk_n)}{|h_n|^2}\frac{1}{L}\bigg)\Bigg]\nonumber\\
&=iK\sum_s s\gamma_s^{a,b}-iKB_4{\rm Im} \bigg(\frac{1}{2\pi}\int_{k=0}^{2\pi}\frac{h^*_k d h_k}{|h_k|^2}\bigg)\nonumber\\
&=iK\sum_s s\gamma_s^{a,b}-iKB_4W_H\nonumber\\
&=iK\sum_s s\gamma_s^{a,b}-iKW_H\sum_s\gamma_s^{a,b}=iK\sum_s(s-W_H)\gamma_s^{a,b},
\end{align}
where $W_H$ is the winding number of the Bloch Hamiltonian
\begin{align}
H(k)=\begin{pmatrix}
    0&h^*_k\\
    h_k&0
     \end{pmatrix},
\end{align}
\begin{align}
W_H=\frac{1}{2\pi i}\int_0^{2\pi} {\rm d} k\frac{{\rm d}}{{\rm d}k}\ln {h}_k=\int_{k=0}^{2\pi}\frac{h^*_k d h_k}{|h_k|^2}.
\end{align}
Finally, we note that $\lambda(K=0^\pm)=\Delta \lambda_0$ corresponds to a non-steady state closed to $\lambda_0=0$, and have ${\rm Re} [\lambda(K=0^\pm)]<0$. Therefore, the winding of the trajectory of $\lambda(K)$ near $K=0$ depends on the sign of $\sum_s(s-W_H)\gamma_s^{a,b}$. Namely, 
  \begin{equation}
    W_0={\rm Sgn}[\sum_s(s-W_H)\gamma_s^{a,b}].  \label{eq:winding_single-tone}
  \end{equation}
  Note that in principle $W_0$ cannot have magnitude larger than one, as it is defined for a negative real reference point which infinitely approaches $\lambda_0=0$ from the negative direction of the real axis, and $\lambda_0=0$ has no degeneracy in our consideration. 

\subsubsection{Non-degeneracy of the steady state}
Numerically we observe that the steady state has no degeneracy, which is important to validate Eq.~\eqref{eq:winding_single-tone}: otherwise the Liouvillian spectrum may winds around $0^-$ multiple times, depending on the spectral trajectory near every degenerate steady states.

In addition to numerical observation, here we provide a proof of the non-degeneracy of the steady state in the single-harmonic limit, i.e., the Hamiltonian contains only a single hopping term $J_{s_0}$ that gives a Fourier harmonic $J_{s_0}e^{-is_0k_n}$ under PBCs.
In this case, the Liouvillian becomes
  \begin{equation}
    \mathcal{L}(K) ={\rm diag}[M_1,M_2,...M_L] + N_L\otimes M_0, 
   \label{eq:LK_JS}
  \end{equation}
  where
  \begin{equation}
    M_n=\begin{pmatrix}
     0 & iJ_{s_0}^*e^{i s_0 k_n'} & -iJ_{s_0}e^{-i s_0k_n} & 0\\
     iJ_{s_0}e^{-is_0k_n’} & -B_4/2 & 0 & -iJ_{s_0}e^{-i s_0k_n}\\
     - iJ_{s_0}^*e^{i s_0k_n} & 0 & -B_4/2 & iJ_{s_0}^*e^{i s_0k_n'}\\
     0 & -iJ_{s_0}^*e^{i s_0k_n} & iJ_{s_0}e^{-i s_0k_n’} & -B_4
     \end{pmatrix},
    \qquad
    M_0=\begin{pmatrix}
    0&0&0&A_{14}\\
    0&0&0&0\\
    0&0&0&0\\
    0&0&0&0
     \end{pmatrix},  
  \end{equation}
  with $A_{14}=\sum_s\gamma_s^{a,b} e^{iKs}/L$ and $B_4=\sum_{s,\alpha}\gamma_s^{\alpha,b}=\sum_{s}\gamma_s^{a,b}$. 
We then consider a  transformation
  \begin{equation}
    \tilde{M}=T_n^{-1} M_n T_n=\begin{pmatrix}
     0 & iJ_{s_0}^* & -iJ_{s_0} & 0\\
     iJ_{s_0} & -B_4/2 & 0 & -iJ_{s_0}\\
     - iJ_{s_0}^*& 0 & -B_4/2 & iJ_{s_0}^*\\
     0 & -iJ_{s_0}^* & iJ_{s_0} & -B_4
       \end{pmatrix},\qquad
   \tilde{M}_0= T_n^{-1} M_0 T_m=e^{-is_0K}M_0,\qquad
T_n=\begin{pmatrix}
     e^{is_0K} & 0 & 0 & 0\\
     0 & e^{-is_0k_n} & 0 & 0\\
     0 & 0 & e^{is_0k_n'} & 0\\
     0 & 0 & 0 & 1
  \end{pmatrix}.
   \label{eq:mappingS}
  \end{equation}
Note that the transformation is equivelant to consider an alternative Fourier transformation,
  \begin{equation}
    |b,l\rangle=\frac{1}{\sqrt{L}}\sum_{k} e^{-ikl}|b,k\rangle,~~~|a,l\rangle=\frac{1}{\sqrt{L}}\sum_{k} e^{-ik(l-s_0)}|a,k\rangle. 
  \end{equation}
The transformed Liouvillian is thus given by
  \begin{equation}
    \tilde{\mathcal{L}}(K) =I_L\otimes \tilde{M} + N_L\otimes \tilde{M}_0,
   \label{eq:LK_simplifiedS}   
  \end{equation}
whose eigenvalues are the same as $\mathcal{L}(K)$.
Since $I_L$ and $N_L$ can be diagonalized and commute with each other, their action on the space spanned by the eigenvectors of $N_L$ reduces $\tilde{\mathcal{L}}(K)$ into a direct sum of several $4\times4$ matrices. That is, as $N_L$ has eigenvalues $L$ (multiplicity 1) and $0$ (multiplicity $L-1$), the spectrum of $\tilde{\mathcal{L}}(K)$ is the union of the following two parts.
\begin{itemize}
 \item [(i)] Eigenvalues of $\tilde{M}$, each with multiplicity $L-1$.
The characteristic polynomial of $\tilde{M}$ is:
  \begin{equation}
    P_{\tilde{M}}= \frac{1}{4}(2\lambda+B_4)^2 \bigl(\lambda^2 + B_4\lambda + 4|J_{s_0}|^2\bigr),
  \end{equation}
which gives the eigenvalues
  \begin{equation}
    \lambda_1 =\lambda_2= -\frac{B_4}{2} ,
   \label{eq:solution1}
  \end{equation}
  \begin{equation}
    \lambda_3 = \frac{-B_4 \pm \sqrt{B_4^2 -16J_{s_0}^2}}{2}.
   \label{eq:solution2}
  \end{equation}
  Each of these eigenvalues appears in $\tilde{\mathcal{L}}(K)$ with multiplicity $(L-1)$.

 \item [(ii)] Eigenvalues of $\tilde{M} +L \tilde{M}_0$, each with multiplicity $1$.

  Characteristic polynomial of $\tilde{M} + L \tilde{M}_0$ is:
  \begin{equation}
   P_{\tilde{M}+L\tilde{M}_0}(\lambda) = \frac{1}{4}(2\lambda+B_4)\, P_3(\lambda),   
  \end{equation}
  where
  \begin{equation}
   P_{3}(\lambda)=2\lambda^3+3B_4\lambda^2+(B_4^2+8|J_{s_0}|^2)\lambda+4|J_{s_0}|^2(B_4-\tilde{A}_{14}L).   
  \end{equation}
$\tilde{A}_{14}=e^{-iKs_0}A_{14}$.
  Thus eigenvalues of $\tilde{M} + L \tilde{M}_0$ include:
  \begin{equation}
    \lambda = -\frac{B_4}{2},
   \label{eq:solution3}
  \end{equation}
  and the three roots of
  \begin{equation}
    \lambda^3 + \frac{3}{2}B_4\lambda^2 + \frac{B_4^2+8|J_{s_0}|^2}{2}\lambda + 2|J_{s_0}|^2(B_4 - \tilde{A}_{14}L) = 0.
   \label{eq:solution4}
  \end{equation}
  Each of these eigenvalues appears in $\tilde{\mathcal{L}}(K)$ with multiplicity $1$.
\end{itemize}
Based on the solutions in Eqs.(\ref{eq:solution1}), (\ref{eq:solution2}), \ref{eq:solution3}, \ref{eq:solution4} and their respective multiplicities, we obtain the spectrum of $\mathcal{L}(k)$ as:
  \begin{equation}
    \lambda = -\frac{B_4}{2}: \quad \text{multiplicity }  (2L - 1),  
  \end{equation}
  \begin{equation}
     \lambda = \frac{-B_4 \pm \sqrt{B_4^2 - 16|J_{s_0}|^2}}{2}: \quad \text{each multiplicity } L-1, 
  \end{equation}
  and the three cubic roots of Eq.~\eqref{eq:solution4}, each appear with multiplicity 1. 
Generally, we assume non-vanishing parameters $J_{s_0}\neq 0$ and $B_4\neq 0$, thus, the only steady state is given by the simple root
\begin{align}
\lambda_0=0
\end{align}
of Eq.~\eqref{eq:solution4} at $K=0$ (so that $B_4 - \tilde{A}_{14}L=0$).

So far we have derived the non-degeneracy of the steady state in the single-harmonic limit. 
Further introducing extra hopping terms generally make the Liouvillian spectrum more complicated, but it is unlikely to introduce a robust degeneracy to the steady states, except for some coincidental ones under some fine-tuned parameters (e.g., balanced dissipation channels with $B_4=0$ in the above single-harmonic limit).

\subsection{Bidirectional dissipation (both $b\rightarrow a$ and $a\rightarrow b$)}
Next, we consider the case with both nonzero $\gamma_s^{a,b}$ and $\gamma_s^{b,a}$ (the subscript $s$ can be different for them).
The Liouvillian is given by
  \begin{equation}
    \mathcal{L}(K) ={\rm diag}[M_1,M_2,...M_L] + N_L\otimes M_0, 
  \end{equation}
  where
  \begin{equation}
    M_n=\begin{pmatrix}
     -B_1 & ih_{n'} & -ih_n^* & 0\\
     ih^*_{n'} & -(B_1+B_{4})/2 & 0 & -ih^*_n\\
     - ih_n & 0 &-(B_1+B_{4})/2  & ih_{n'}\\
     0 & -ih_n & ih^*_{n'}& -B_{4}
     \end{pmatrix},
    \qquad
    M_0=\begin{pmatrix}
    0&0&0&A_{14}\\
    0&0&0&0\\
    0&0&0&0\\
    A_{41}&0&0&0
     \end{pmatrix},  
  \end{equation}
with $A_{14}=\sum_s\gamma_s^{a,b} e^{iKs}/L$, $A_{41}=\sum_s\gamma_s^{b,a} e^{iKs}/L$,  $B_1=\sum_{s}\gamma_s^{b,a}$, and $B_4=\sum_{s}\gamma_s^{a,b}$. 

\subsubsection{Steady state at $K=0$ (single-harmonic limit)}
Similarly, since $k_n=k'_n$ when $K=0$,
we introduce auxiliary matrices
\begin{equation}
    M_{n,{\rm aux}}=M_n+LM_0=\begin{pmatrix}
     -B_1 & ih_{n'} & -ih_n^* & B_4\\
     ih^*_{n'} & -(B_1+B_{4})/2 & 0 & -ih^*_n\\
     - ih_n & 0 &-(B_1+B_{4})/2  & ih_{n'}\\
     B_1 &  -ih_n & ih^*_{n'}& -B_4
     \end{pmatrix},
\qquad\mathcal{L}_{\rm aux}={\rm diag}[M_{1,{\rm aux}},M_{2,{\rm aux}},...,M_{L,{\rm aux}}],
  \end{equation}
with $H_n=-i\sum_s J_s e^{-is k_n}$.
$M_{n,{\rm aux}}$ contains a simple root $\lambda_{\rm aux}=0$, with its left eigenvector $\langle\langle \psi'_n|$ given by
\begin{align}
\langle\langle \psi'_n|M_{n,{\rm aux}}|=0,\qquad| \psi'_n\rangle\rangle=
\begin{pmatrix}
\psi'_{n,1}\\
\psi'_{n,2}\\
\psi'_{n,3}\\
\psi'_{n,4}
\end{pmatrix}=
\begin{pmatrix}
1\\
0\\
0\\
1
     \end{pmatrix}
\qquad\Rightarrow\qquad
\langle\langle \Psi'|\mathcal{L}_{\rm aux}|=0,
\qquad
|\Psi'\rangle\rangle=\frac{1}{\sqrt{L}}\begin{pmatrix}
|\psi'_1\rangle\rangle\\
|\psi'_2\rangle\rangle\\
...\\
|\psi'_L\rangle\rangle
\end{pmatrix}.
\end{align}
Next, we act the original Liouvillian at $K=0$ on $\langle\langle \Psi'|$,
\begin{align}
\langle\langle \Psi'|\mathcal{L}(0)=\langle\langle \Phi'|,\qquad |\Phi'\rangle\rangle=\frac{1}{\sqrt{L}}\begin{pmatrix}
|\phi_1\rangle\rangle\\
|\phi_2\rangle\rangle\\
...\\
|\phi_L\rangle\rangle
\end{pmatrix},\qquad|\psi_n\rangle\rangle=\begin{pmatrix}
\phi_{n,1}\\
\phi_{n,2}\\
\phi_{n,3}\\
\phi_{n,4}
\end{pmatrix}.
\end{align}
Since $\mathcal{L}(0)$ and $\mathcal{L}_{\rm aux}$  are different only in their $(4n-3,4m)$ and $(4n,4m-3)$ elements with $n,m\in\{1,2,...,L\}$, we obtain
\begin{align}
\langle\langle \Phi'|&=\langle\langle \Psi'| \mathcal{L}(0)=\langle\langle \Psi'|\big[\mathcal{L}(0)-\mathcal{L}_{\rm aux}\big]
=\frac{1}{\sqrt{L}}[(A_{41}L-B_1),0,0,(A_{14}L-B_4),(A_{41}L-B_1),0,0,(A_{14}L-B_4),...]=0
\end{align}
since $A^*_{14}L=B_4$ and $A^*_{41}L=B_1$ when $K=0$.
Thus we obtain $\langle\langle\Psi'|\mathcal{L}(K=0)|$=0, meaning that $|\Psi'\rangle\rangle$ is the left-eigenvector of the steady state with $\lambda_0=0$.

\bigskip 

In contrast to the left-eigenvector, the right-eigenvector of the steady state takes a more complicated form.
Starting from the auxiliary matrices, we obtain
\begin{align}
M_{n,{\rm aux}}|\psi_n\rangle\rangle=0,\qquad|\psi_n\rangle\rangle=
\begin{pmatrix}
\psi_{n,1}\\
\psi_{n,2}\\
\psi_{n,3}\\
\psi_{n,4}
\end{pmatrix}=
\begin{pmatrix}
{\big[B_4^2+B_1B_4+4|h_n|^2\big]}/{\big[B_1^2+B_1B_4+4|h_n|^2\big]}\\
{-2i(B_1-B_4)h^*_n}/{\bigg[ B_1^2+B_1B_4+4|h_n|^2\bigg]}\\
{2i(B_1-B_4)h_n}/{\bigg[ B_1^2+B_1B_4+4|h_n|^2\bigg]}\\
1
     \end{pmatrix}
~~\Rightarrow~~
\mathcal{L}_{\rm aux}|\Psi\rangle\rangle=0,
\qquad
|\Psi\rangle\rangle=\frac{1}{\sqrt{L}}\begin{pmatrix}
|\psi_1\rangle\rangle\\
|\psi_2\rangle\rangle\\
...\\
|\psi_L\rangle\rangle
\end{pmatrix};
\end{align}
\begin{align}
|\Phi\rangle\rangle&=\mathcal{L}(0)|\Psi\rangle\rangle=\big[\mathcal{L}(0)-\mathcal{L}_{\rm aux}\big]|\Psi\rangle\rangle\nonumber\\
&=\frac{1}{\sqrt{L}}[(A_{14}L-B_4),0,0,\left(A_{41}\sum_n\frac{B_4^2+B_1B_4+4|h_n|^2}{B_1^2+B_1B_4+4|h_n|^2}-B_1\frac{B_4^2+B_1B_4+4|h_1|^2}{B_1^2+B_1B_4+4|h_1|^2}\right),\nonumber\\
&~~~~~~(A_{14}L-B_4),0,0,\left(A_{41}\sum_n\frac{B_4^2+B_1B_4+4|h_n|^2}{B_1^2+B_1B_4+4|h_n|^2}-B_1\frac{B_4^2+B_1B_4+4|h_2|^2}{B_1^2+B_1B_4+4|h_2|^2}\right),...]^T\nonumber\\
&\Rightarrow \phi_{n,2}=\phi_{n,3}=0,\qquad \phi_{n,1}=A_{14}L-B_4,\qquad \phi_{n,4}=\left(A_{41}\sum_{n'}\frac{B_4^2+B_1B_4+4|h_{n'}|^2}{B_1^2+B_1B_4+4|h_{n'}|^2}-B_1\frac{B_4^2+B_1B_4+4|h_n|^2}{B_1^2+B_1B_4+4|h_n|^2}\right).
\end{align}
We have $\phi_{n,1}=0$ when $K=0$;
however, $\phi_{n,4}=0$ further requires $|h_n|=|h_{n'}|$ for arbitrary $n$ and $n'$.
This condition can be satisfied under the single-harmonic limit of the Hamiltonian, i.e., the Hamiltonian contains only a single hopping term, corresponding to $h_n=J^*_{s_0}e^{is_0 k_n}$.
In other words, $|\Psi\rangle\rangle$ is the right-eigenvector of the steady state in the single-harmonic limit.

\subsubsection{Deviating from $K=0$ (single-harmonic limit)}
For a small but nonzero $K=k'-k$,
The Liouvillian in the single-harmonic limit can be written as
\begin{align}
\mathcal{L}(K)=\mathcal{L}(0)+\Delta\mathcal{L},\qquad
\Delta \mathcal{L}={\rm diag}[\Delta M_1,\Delta M_2,...,\Delta M_L]+N_L\otimes \Delta M_0,
\end{align}

 \begin{equation}
    \Delta M_n(K)=\begin{pmatrix}
     0 & -KJ^*_{s_0} {s_0} e^{i{s_0}k_n}& 0 & 0\\
     KJ_{s_0} {s_0} e^{-i{s_0}k_n}& 0 & 0 & 0\\
     0 & 0 & 0 & -KJ^*_{s_0} {s_0} e^{i{s_0}k_n}\\
     0 & 0 & KJ_{s_0} {s_0} e^{-i{s_0}k_n} & 0
     \end{pmatrix},
    \qquad
    \Delta M_0(K)=\begin{pmatrix}
    0&0&0&iK\sum_s s\gamma_s^{a,b}/L\\
    0&0&0&0\\
    0&0&0&0\\
    iK\sum_s s\gamma_s^{b,a}/L&0&0&0
     \end{pmatrix}.
  \end{equation}
The first-order perturbation to the steady-state eigenvalue $\lambda_0=0$ is given by
\begin{align}
&\Delta \lambda_0\propto\langle\langle \Psi'| \Delta\mathcal{L}|\Psi\rangle\rangle=i\frac{K}{L}\sum_n \Bigg(
\sum_s s\gamma_s^{a,b}+\frac{B_4^2+B_1B_4+4|J_{s_0}|^2}{B_1^2+B_1B_4+4|J_{s_0}|^2}\sum_s s\gamma_s^{b,a}
+\frac{4{s_0}(B_1-B_4)|J_{s_0}|^2}{B_1^2+B_1B_4+4|J_{s_0}|^2}\Bigg)\nonumber\\
&\propto iK\left[
\left(B_1^2+B_1B_4+4|J_{s_0}|^2\right)\sum_s s\gamma_s^{a,b}+\left(B_4^2+B_1B_4+4|J_{s_0}|^2\right)\sum_s s\gamma_s^{b,a}
+{4{s_0}(B_1-B_4)|J_{s_0}|^2}\right] \nonumber\\
&= iK\left[
\left(B_1^2+B_1B_4\right)\sum_s s\gamma_s^{a,b}+\left(B_4^2+B_1B_4\right)\sum_s s\gamma_s^{b,a}
+4|J_{s_0}|^2\sum_s\left( [s-s_0]\gamma_s^{a,b}+[s+s_0]\gamma_s^{b,a}\right)\right].
\label{eq:perturbed}
\end{align} 
Since $\Delta \lambda_0$ gives the eigenvalue deviated from the steady state at $\lambda_0=0$, the Liouvillian winding number is given by
\begin{align}
\label{eq:s77}
W_0={\rm Sgn}[\Delta \lambda_0]={\rm Sgn}\left[
\left(B_1^2+B_1B_4\right)\sum_s s\gamma_s^{a,b}+\left(B_4^2+B_1B_4\right)\sum_s s\gamma_s^{b,a}
+4|J_{s_0}|^2\sum_s\left( [s-s_0]\gamma_s^{a,b}+[s+s_0]\gamma_s^{b,a}\right)\right].
\end{align}
We can see that even in this simplified single-harmonic case, the Liouvillian winding number $W_0$ depends not only on the Hamiltonian winding  number $W_H=s_0$, but also on the magnitude of the hopping parameter $J_s$.
In other words, the Liouvillian spectral winding is associated with the Hamiltonian not only via its topology, but also the explicit parameters.

\begin{figure}
    \centering
    \includegraphics[width=0.5\linewidth]{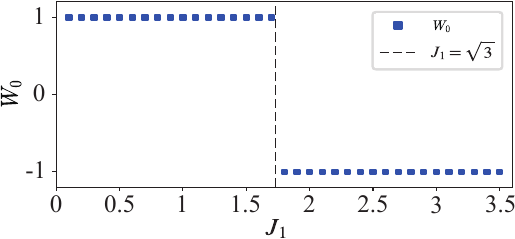}
    \caption{Liouvillian spectrum winding number $W_0$ for the single-harmonic system with $J_s=0$ for all $s\neq 1$.
    The dissipation parameters are $\gamma_0^{a,b}=3$ and $\gamma_1^{b,a}=1$.
    Substituting our parameters into Eq.~\eqref{eq:s77}, we obtain $W_0={\rm sgn}[12-4J_1^2]$, which yields a phase transition point at $J_1=\sqrt{3}$ (black dashed line in the figure).
    }
    \label{fig:S1}
\end{figure}

{Fig.~\ref{fig:S1}
shows the numerical results 
of a single-harmonic example with $s_0=1$ ($J_s=0$ for $s\neq1$), with sublattice-bidirectional dissipation given by $\gamma_0^{a,b}$ and $\gamma_1^{b,a}$.
It can be seen that the topology of the Liouvillian undergoes a phase transition 
as the strength of $J_1$ varies, as predicted by Eq.~\eqref{eq:s77}. 
Meanwhile, the Hamiltonian topology remains unchanged as the band-topological winding number $W_H=s_0$ in the single-harmonic limit.
}

\section{Coherent LSE with band-topological steady state}\label{Sec:band-topo}
In this section, we prove that the topological edge state of the Hamiltonian gives the left-LSE steady state in our example model.

\subsection{Topological edge states of the Hamiltonian}
The Hamiltonian is given by 
\begin{align}
    \hat{H}_{\rm SSH}=\sum_{l=1}^L\left(J_0|a,l\rangle\langle b,l|+J_1|a,l+1\rangle\langle b,l|\right)+{\rm h.c}..
\end{align}
For an eigenstate $|\Psi\rangle=\sum_{l}(\psi_{a,l}|a,l\rangle+\psi_{b,l}|b,l\rangle)$ with eigenenergy $E$, the Schr\"odinger equation yields
\begin{align}
   & J_0\psi_{b,l}+J_1\psi_{b,l-1}=E\psi_{a,l},\quad
    J^*_0\psi_{a,l}+J^*_1\psi_{a,l+1}=E\psi_{b,l},
    \\
    \Rightarrow \quad &J_0\psi_{b,l}+J_1\psi_{b,l-1}=0,\quad
    J^*_0\psi_{a,l}+J^*_1\psi_{a,l+1}=0,\quad l=1,2,...L\label{eq:recurrence}
\end{align}
for topological edge states at $E=0$,
with open boundary conditions $\psi_{\alpha,0}=\psi_{\alpha,L+1}=0$.
Thus, we obtain
\begin{align}
    &\psi_{a,l}\propto \left(-\frac{J_0^*}{J_1^*}\right)^{l-1},\quad\psi_{b,l}=0~{\rm for~the~left-edge~state~}|\Psi_{\rm left}\rangle,\label{eq:T_edge_L}\\
    &\psi_{b,l}\propto \left(-\frac{J_0}{J_1}\right)^{L-l},\quad\psi_{a,l}=0~{\rm for~the~right-edge~state~}|\Psi_{\rm right}\rangle\label{eq:T_edge_R}.
\end{align}

\subsection{Band-topological steady state of the Liouvillian}
Following the discussion in previous sections and in the main text, we consider $b\rightarrow a$ dissipation described by $\hat{D}_{l,s}^{a,b}=\sqrt{\gamma_s^{a,b}}|a,l+s\rangle\langle b,l|$.
Apparently, such dissipation does not act on the left-edge state that is polarized within sublattice $a$.
Explicitly,
\begin{align}
&\rho_{\rm left}=|\Psi_{\rm left}\rangle\langle\Psi_{\rm left}|\propto\sum_{l,l'}\left(-\frac{J_0^*}{J_1^*}\right)^{l+l'-2}|a,l\rangle\langle a,l'|,\\
&\hat{D}_{l_1,s}^{a,b}\hat{\rho}_L[\hat{D}_{l_1,s}^{a,b}]^\dagger=\gamma_s^{a,b}\sum_{l,l'}\left(-\frac{J_0^*}{J_1^*}\right)^{l+l'-2}|a,l_1+s\rangle\langle b,l_1|a,l\rangle\langle a,l'|b,l_1\rangle\langle a,l_1+s|=0,\\
&\{[\hat{D}_{l_1,s}^{a,b}]^\dagger\hat{D}_{l_1,s}^{a,b},\hat{\rho}_L\}=\left\{\gamma_s^{a,b}|b,l_1\rangle\langle b,l_1|,\quad\sum_{l,l'}\left(-\frac{J_0^*}{J_1^*}\right)^{l+l'-2}|a,l\rangle\langle a,l'|\right\}=0.
\end{align}
Thus the steady state is a dark state given by $|\Psi_{\rm left}\rangle$.
However, when the Hamiltonian is topologically trivial (with $W_0=0$), all eigenstates are bulk states without sublattice-polarization. Thus the right-LSE state in Fig. 2(e) of the main text has to be mixed state.

\subsection{Finite-size effect of the bulk-boundary correspondence of the spectral winding topology}

For the SSH model under a finite size, 
$|\Psi_{\rm left}\rangle$ and $|\Psi_{\rm right}\rangle$ are not the exact eigenstates of the Hamiltonian $\hat{H}_{\rm SSH}$.
Instead, topological edge states are given by the linear combination of them, with small but nonzero size-dependent energies $\epsilon\sim (J_0/J_1)^L$.
Explicitly, a finite-size system has
\begin{align}
    \hat{H}_{\rm SSH}|\Psi_{\rm left/right}\rangle=\epsilon|\Psi_{\rm rignt/left}\rangle.
\end{align}
Nevertheless, In this case, 
$|\Psi_{\rm left}\rangle$  still provides a good approximation to the steady state when the finite-size energy splitting ($\epsilon$) remains sufficiently small.
As the system approaches the topological phase boundary ($J_0=J_1$), the edge-state localization length increases and $\epsilon$ becomes appreciable, thereby weakening the dark-state description.
Consequently, a sufficiently strong rightward dissipation $\gamma_1^{a,b}$ can drive the steady state toward the right boundary even within the topologically nontrivial regime $J_0<J_1$, as shown in Fig. 3 of the main text.
In particular, Fig. 3(d) clearly indicates that this inconsistency between $W_0$ and the localization direction of the steady state is a finite-size effect.


On the other hand, we note that
an exact bulk-boundary correspondence of the LSE can also be recovered in finite-size systems, by removing one lattice site from the original OBC lattice.
For example, if we remove the right-most lattice site $(b,L)$, the system supports only one topological edge state either when $W_H=1$ or $W_H=0$ [the later case becomes topologically nontrivial when redefining the unit cells with sites $(b,l)$ and $(a,l+1)$].
The wavefunction satisfies the same recurrence relation of Eq.~\eqref{eq:recurrence}, but with different numbers of equations for the two sublattices,
\begin{align}
    &J_0\psi_{b,l}+J_1\psi_{b,l-1}=0,\quad l=1,2,...,L\nonumber\\
    &J^*_0\psi_{a,l}+J^*_1\psi_{a,l+1}=0,\quad l=1,2,...,L-1;
\end{align}
and the boundary conditions changed to $\psi_{\alpha,0}=\psi_{b,L}=\psi_{a,L+1}=0$.
Thus the sublattice-polarized state
\begin{align}
|\Psi_{\rm left}\rangle=\sum_{l=1}^L\psi_{a,l}|a,l\rangle,\quad \psi_{a,l}\propto \left(-\frac{J_0^*}{J_1^*}\right)^{l-1}
\end{align}
is always an eigenstate of the Hamiltonian, and hence gives the dark steady state of the open quantum system.

\section{Competition between the coherent LSE and incoherent LSE}\label{sec:compete}

\begin{figure}
    \centering
    \includegraphics[width=1\linewidth]{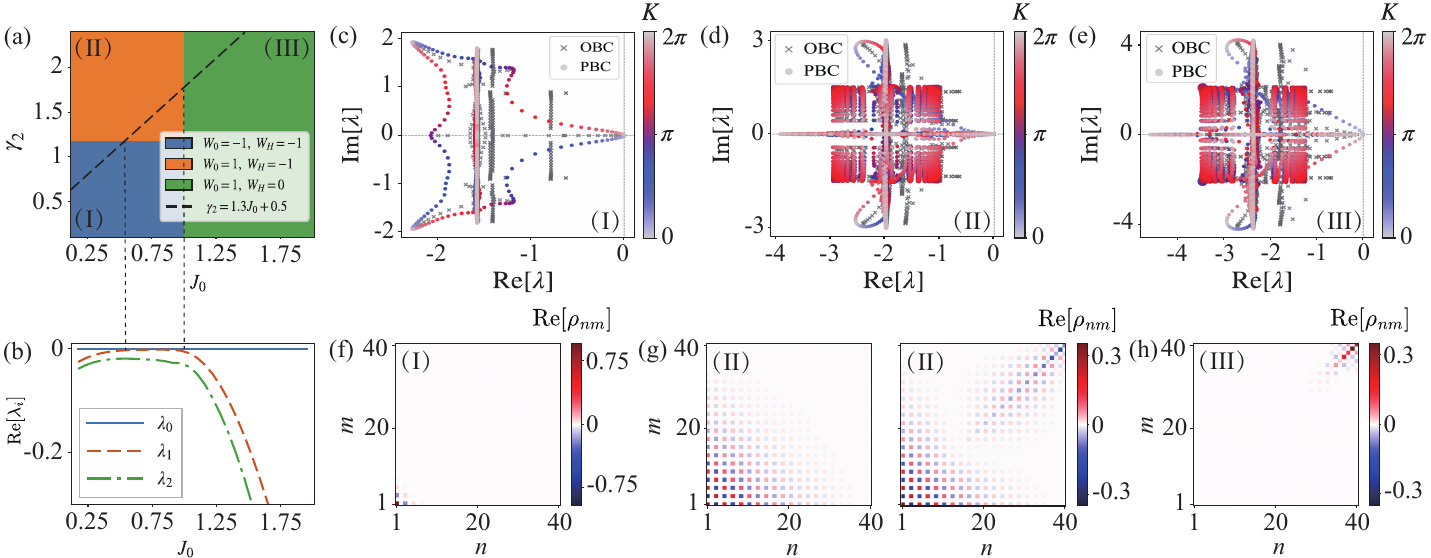}
    \caption{(a) Phase diagram of the SSH model with $\gamma_{0,1,2}^{a,b}\equiv\gamma_{0,1,2}$ dissipation. Different colors represent different phases, with each phase characterized by distinct values of $W_H$ or $W_0$. 
    (b) Real parts of the three OBC eigenvalues $\lambda_i$ with the lowest negative ${\rm Re}[\lambda_i]$, for the parameters along the dash line in (a).
    (c), (d), and (e) show the Liouvillian spectra under OBC (gray) and PBC (colored) for the three phases in (a), respectively (at parameter points taken on the dashed line). Colors represent different $K$ values of the PBC eigenvalues. The specific parameters are $J_0=0.2$, $0.8$, and $1.4$, respectively, with $\gamma_2$ determined by $\gamma_2=1.3J_0+0.5$.
    (f), (g), and (h) show the density matrices corresponding to the zero (and near-zero) eigenvalues under OBC for the cases in (c), (d), and (e), respectively.
    Other parameters are  $J_1=1$, $\gamma_0=\gamma_1=1.2$, and $L=20$.
    }
    \label{fig:s3}
\end{figure}

As an example, we consider a dissipative extension of the Su-Schrieffer-Heeger (SSH) model,
given by the Hamiltonian of Eq.(1) in the main text with nonzero $J_0$ and $J_1$ only (where $W_H$ takes only $0$ or $1$), and three dissipative processes $\gamma_s^{a,b}\equiv\gamma_s$ with $s=0,1,2$.
The Liouvillian winding number is reduced to
\begin{align}
W_0&={\rm Sgn}[(2-W_H)\gamma_2+(1-W_H)\gamma_1-W_H\gamma_0]\nonumber=\begin{cases}
 1, & \text{if~}W_H=0 \\
 {\rm Sgn}[\gamma_2-\gamma_1], & \text{if~}W_H=1
\end{cases}.
\label{eq:W_SSH}
\end{align}
In Fig.~\ref{fig:s3}(a), we show numerical results of two topological invariants in the parameter space of $J_0$ and $\gamma_2$, which verify the above relation.
Figs. \ref{fig:s3}(c) to \ref{fig:s3}(e) provide illustrative cases of the full Liouvillian spectra for the three different phases.
It is clearly visible that the PBC spectrum forms a loop encircling the OBC one, whose winding direction is consistent with $W_0$.
The system under OBCs also shows different behaviors in the three regimes.
Similar to the model discussed in the main text, the localization directions consists with the prediction of $W_0$ in Phases I and III.
However, phase II supports two nearly-zero-eigenvalue modes, which together show a bipolar localization of the steady state.

Physically, the inconsistency between $W_0$ and the steady-state localization can be interpreted from the competition between two edge-localization mechanisms in our system:
 (i) dissipation-induced LSE characterized by $W_0$, and (ii) band-topological localization from the Hamiltonian with a nonzero $W_H$, which gives a dark state to the system (see Sec. \ref{Sec:band-topo}).
 The two mechanisms together generate the coherent LSE in phase I [Fig.~\ref{fig:s3}(f)], and the incoherent LSE in phase III is induced solely by the former~[Fig.~\ref{fig:s3}(h)].
However, they lead to opposite localization directions in phase II.
The nearly-zero-eigenvalue modes arise from their competition, thus exhibiting
coherent (incoherent) behaviors on the left-hand (right-hand) side of the lattice [Fig.~\ref{fig:s3}(g)].
In addition, we note that these two modes
have exponentially different eigenvalues: $\lambda_0\sim 10^{-15}$ and $\lambda_1\sim10^{-3}$ for the chosen parameters. 
In other words, they represent a steady state and a metastable mode of the system, respectively.
Nevertheless, the bulk-boundary correspondence of the Liouvillian spectral winding topology still holds:
$W_0=0$ predicts a rightward localization of eigenmodes of the Liouvillian, which is mixed with the dark state arising from the Hamiltonian band-topology.

\section{Chiral-symmetric multi-band systems}
\subsection{Model and topological invariants}
  In the main text, We consider generic one-dimensional two-band models with chiral symmetry, described by the Hamiltonian
\begin{equation}
    \hat{H}=\sum_{l=1}^L\sum_{s}J_{s} |a,l+s\rangle\langle b,l|+{\rm h.c.},
   \label{eq:H}    
\end{equation}
where $l$ labels the unit cell, $L$ is the system size,
and $|s|<L$ denotes the range of hopping from pseudospin or sublattice $b$ to $a$. 
In this section, we consider the following multi-band generalization:
\begin{align}    \hat{H}=\sum_{l=1}^L\sum_{s}\sum_{\mu}T_{\mu}^{(s)} |a,\mu,l+s\rangle\langle b,\mu,l|+{\rm h.c.}.
  \label{eq:H_extended}
\end{align}
Here, each sublattice ($a$ or $b$) is assumed to have a $N_\mu$-dimensional internal degree of freedom, labeled by $\mu=1,2,...,N_\mu$, and
$T^{s}$ is an $N_\mu\times N_\mu$ matrix describing the $s$-th nearest-neighbor hopping.

The quantum jump operators in the multi-band system are given by
\begin{align}
   [\hat{D}_{l,s}^{a,a}]_{\mu\mu'} = \sqrt{\gamma_{\mu\mu'}^{(s;a,a)}} |a, \mu, l + s\rangle \langle a, \mu', l|,
   \quad
   [\hat{D}_{l,s}^{b,b}]_{\mu\mu'} = \sqrt{\gamma_{\mu\mu'}^{(s;b,b)}} |b, \mu, l + s\rangle \langle b, \mu', l|,
  \\
   [\hat{D}_{l,s}^{a,b}]_{\mu\mu'} = \sqrt{\gamma_{\mu\mu'}^{(s;a,b)}} |a, \mu, l + s\rangle \langle b, \mu', l|,
   \quad
   [\hat{D}_{l,s}^{b,a}]_{\mu\mu'} = \sqrt{\gamma_{\mu\mu'}^{(s;b,a)}} |b, \mu, l + s\rangle \langle a, \mu', l|.
   \end{align}

To characterize the topological properties, we take the Fourier transformation
\begin{equation}
  |a,\mu,l\rangle=\frac{1}{\sqrt{L}}\sum_k e^{-ikl}|a,\mu,k\rangle,
\end{equation}
then we get
\begin{equation}
  H(k)=\begin{pmatrix}
     0& X^\dag_k\\
     X_k & 0
    \end{pmatrix},  
    \label{eq:Hk_extended}
\end{equation}
with $X_k=\sum_s [T^{(s)}]^\dag e^{iks}= {\rm diag}\big[x_{k,1},x_{k,2},...,x_{k,N_\mu}]$ also a diagonal matrix, $x_{k,\mu}=\sum_s [T_\mu^{(s)}]^* e^{iks}$.
The topological invariants can be defined as
\begin{equation}
   W_H=\frac{1}{2\pi i}\int_{-\pi}^{\pi}dk \partial_k \ln \det X_k= \sum_\mu\frac{1}{2\pi i}\int_{-\pi}^{\pi}dk \partial_k \ln x_{k,\mu}. 
\end{equation}
The Liouvillian spectral winding number $W_0$ is defined in the same way as in Eq. (7) of the main text.

\subsection{A numerical example}
Here we take a $2\times2$ band model as an example. 
We write the hopping matrices as
\begin{equation}
  T_\mu^{(0)}\to T^{(0)}= 
  \begin{pmatrix}
     v& \delta_1\\
     \delta_1 & v
    \end{pmatrix},~~~~~~~
  T_\mu^{(1)}\to T^{(1)}= 
  \begin{pmatrix}
     w& \delta_2\\
     \delta_2 & w
    \end{pmatrix}, \label{eq:Mband_hopping}
\end{equation}
and consider sublattice-unidirectional chiral-symmetric dissipation
\begin{equation}
  \gamma_{\mu\mu'}^{(0;a,b)}\to \gamma^{(0)},~~~~~
  \gamma_{\mu\mu'}^{(1;a,b)}\to \gamma^{(1)},\label{eq:Mband_dissipation}
\end{equation}
with all other dissipation coefficients set to zero.

\begin{figure}
    \centering
    \includegraphics[width=1\linewidth]{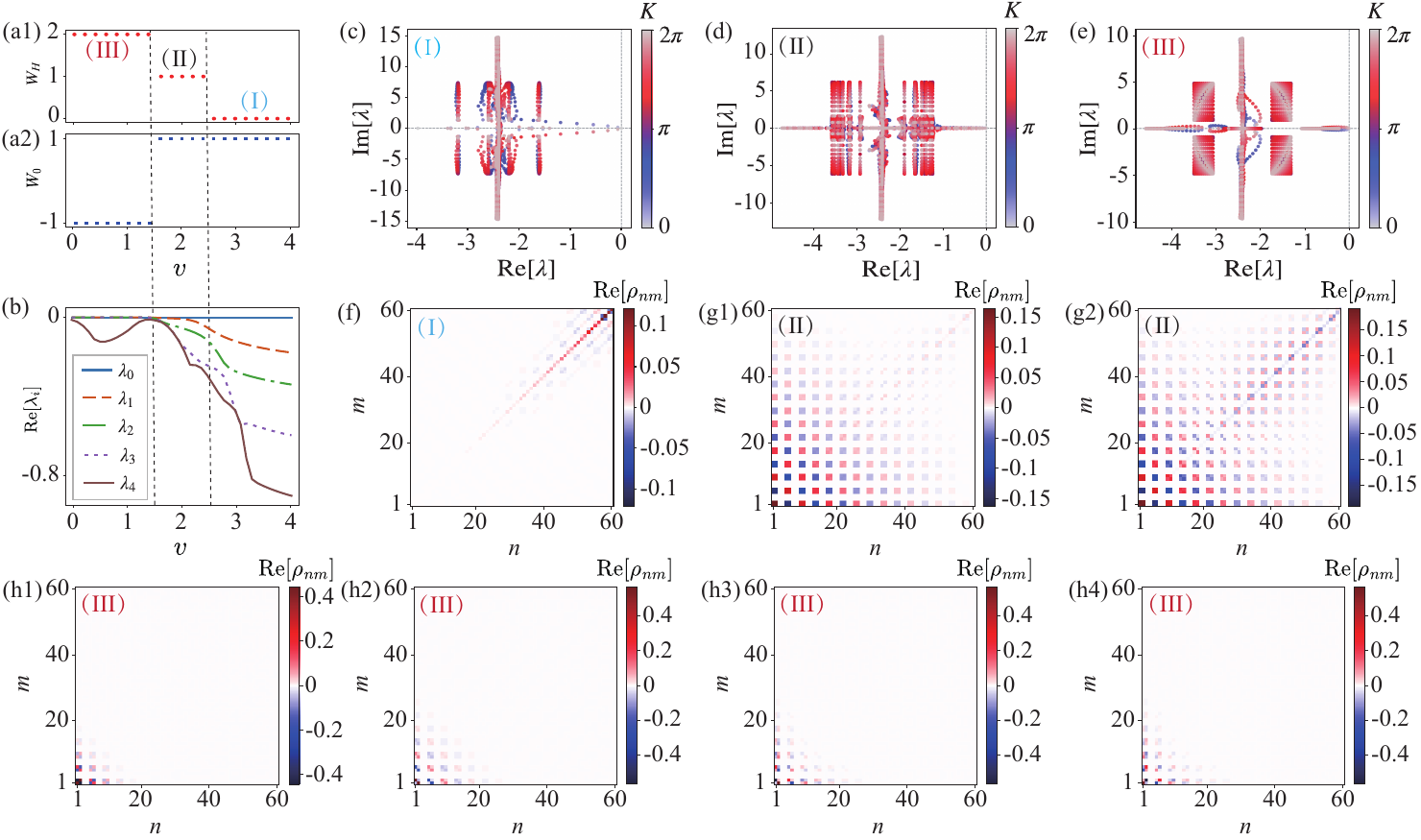}
    \caption{Numerical results of the multi-band model
    of Eqs.~\eqref{eq:Mband_hopping} and~\eqref{eq:Mband_dissipation}.
    (a1) the band winding number $W_H$ of the Hamiltonian and (a2) the Liouvillian spectral winding number $W_0$, as functions of $v$.
    (b) Real parts of the five OBC eigenvalues $\lambda_i$ with the lowest negative ${\rm Re}[\lambda_i]$.
    (c), (d), and (e) show the Liouvillian spectra under OBC (gray) and PBC (colored) for the three phases in (a), respectively. Colors represent different $K$ values of the PBC eigenvalues. The specific parameters are $v=1$, $1.8$, and $3.4$, respectively.
    (f), (g), and (h) show the density matrices corresponding to the zero (and near-zero) eigenvalues under OBC for the cases in (c), (d), and (e), respectively.
    The other parameters are $w=2$, $\delta_1=0.75$, $\delta_2=1.25$, $\gamma^{(0)}=\gamma^{(1)}=1.2$, and $L=15$.
     }
    \label{fig:s4}
\end{figure}

Fig.~\ref{fig:s4}(a) demonstrate the two topological invariants of this model, which unveil three topologically different phases.
In Fig.~\ref{fig:s4}(b) we illustrate the five eigenvalues with the lowest negative real parts, which show different (near-)degeneracy in the three phases.
Typical Liouvillian spectra of the three phases are displayed in Figs.~\ref{fig:s4}(c) to ~\ref{fig:s4}(e).
In phase I, $W_H=0$ indicates the absence of band-topological edge states, and $W_0=1$ predicts a rightward LSE, leading to an incoherent steady state localized at the right edge [Fig. \ref{fig:s4}(f)].
In phase II, $W_H=1$ predicts a left-edge state of the Hamiltonian, which gives a dark state of the open system; and $W_0=1$ predicts a rightward LSE.
Consequently, their competition lead to two nearly-degenerate modes with a bipolar distribution.
These two phases are qualitatively the same as the two-band model discussed in Sec.~\ref{sec:compete} [see Figs.~\ref{fig:s3}(g) and \ref{fig:s3}(h)].

Phase III is slightly different: $W_H=2$ predicts two left-edge states of the Hamiltonian, and $W_0=-1$ predicts a leftward LSE.
For the sublattice-unidirectional dissipation we consider, the two left-edge states become two dark states, which are expected to give from zero-eigenvalue modes of the Liouvillian (two steady states and their coherences).
Indeed, we observe a four-fold degeneracy in Fig. \ref{fig:s4}(b),  and the corresponding eigenmodes are all localized toward the left edge [Fig.~\ref{fig:s4}(h)].

\section{Chiral-asymmetric quantum jumps}
In this section, we derive the Liouvillian spectral winding number under chiral-asymmetric quantum jumps.
The asymmetric condition is given by [in contrast to Eq.~\eqref{supp_eq:chiral}]
\begin{align}
\sigma_z\hat{D}_p\sigma_z=\hat{D}_p,
\end{align}
which also represents a unitary symmetry and allows quantum jumps between the same sublattices (i.e., $\gamma_s^{a,a}$ and $\gamma_s^{a,a}$).

The Liouvillian superoperator in $K$-space becomes
  \begin{equation}
    \mathcal{L}(K)={\rm diag}[M_1,M_2,...M_L]+N_L\otimes M_0,
    \label{eq:LKS}
  \end{equation}
  with $N_L$ an $L\times L$ matrix with all elements being $1$, and 
  \begin{equation}
    M_n=\begin{pmatrix}
     -B_{1} & ih_{n'} & -ih_n^* & 0\\
     ih^*_{n'} & -(B_{1}+B_{4})/2 & 0 & -ih^*_n\\
     - ih_n & 0 &-(B_{1}+B_{4})/2  & ih_{n'}\\
     0 & -ih_n & ih^*_{n'}& -B_{4}
     \end{pmatrix},
    ~~M_0=\begin{pmatrix}
     A_{11} &0 & 0 & 0\\
     0 & 0& 0 & 0\\
     0 & 0 & 0 & 0\\
     0 & 0& 0& A_{44}
    \end{pmatrix}.
   \label{eq:MNM0}
  \end{equation}
  The coefficients in Eq.~(\ref{eq:MNM0}) are
  \begin{equation}
   \begin{aligned}
    &A_{11}=\sum_s\gamma_s^{a,a} e^{iKs}/L,~~
    A_{44}=\sum_s\gamma_s^{b,b} e^{iKs}/L,\\
    &B_{1}=\sum_{s}\gamma_s^{a,a},~~
    B_{4}=\sum_{s}\gamma_s^{b,b},~~
    h_n=h(k_n)=\sum_s J^*_s e^{is k_n},
   \end{aligned}
  \end{equation}
  with $k_n=2n\pi/L$ and  $k_n'=K+k_n=K+2n\pi/L$. Similar to the chiral-symmetric case, we consider auxiliary matrices
\begin{equation}
    M_{n,{\rm aux}}=M_n+LM_0=
\begin{pmatrix}
     0 & ih_n & -ih^*_n & 0\\
     ih^*_n & -(B_{1}+B_{4})/2& 0 & -ih^*_n\\
     -ih_n & 0 & -(B_{1}+B_{4})/2 & ih_n\\
     0 &  -ih_n & ih^*_n& 0
     \end{pmatrix},
\qquad\mathcal{L}_{\rm aux}={\rm diag}[M_{1,{\rm aux}},M_{2,{\rm aux}},...,M_{L,{\rm aux}}],
  \end{equation}
with 
$h_n=h(k_n)=\sum_s J^*_s e^{is k_n}$.
$M_{n,{\rm aux}}$ contains a simple root $\lambda_{\rm aux}=0$,  
\begin{align}
M_{n,{\rm aux}}|\psi_n\rangle\rangle=0,\qquad|\psi_n\rangle\rangle=
\begin{pmatrix}
\psi_{n,1}\\
\psi_{n,2}\\
\psi_{n,3}\\
\psi_{n,4}
\end{pmatrix}=
\begin{pmatrix}
1\\
0\\
0\\
1
     \end{pmatrix}
\qquad\Rightarrow\qquad
\mathcal{L}_{\rm aux}|\Psi\rangle\rangle=0,
\qquad
|\Psi\rangle=\frac{1}{\sqrt{L}}\begin{pmatrix}
|\psi_1\rangle\rangle\\
|\psi_2\rangle\rangle\\
...\\
|\psi_L\rangle\rangle
\end{pmatrix}.
\end{align}
Next, we act the original Liouvillian at $K=0$ on the column vector $|\Psi\rangle\rangle$,
\begin{align}
\mathcal{L}(0)|\Psi\rangle\rangle=|\Phi\rangle\rangle=\frac{1}{\sqrt{L}}\begin{pmatrix}
|\phi_1\rangle\rangle\\
|\phi_2\rangle\rangle\\
...\\
|\phi_L\rangle\rangle
\end{pmatrix},\qquad|\phi_n\rangle\rangle=\begin{pmatrix}
\phi_{n,1}\\
\phi_{n,2}\\
\phi_{n,3}\\
\phi_{n,4}
\end{pmatrix}.
\end{align}
$\mathcal{L}(0)$ and $\mathcal{L}_{\rm aux}$  are different only in their $(4n-3,4m-3)$ and $(4n,4m)$ elements with $n,m\in\{1,2,...,L\}$, thus we have
\begin{align}
|\Phi\rangle\rangle&=\mathcal{L}(0)|\Psi\rangle\rangle=\big[\mathcal{L}(0)-\mathcal{L}_{\rm aux}\big]|\Psi\rangle\rangle
=\frac{1}{\sqrt{L}}[(A_{11}L-B_1),0,0,(A_{44}L-B_4),(A_{11}L-B_1),0,0,(A_{44}L-B_4),...]^T=0,
\end{align}
as $A_{11}L=B_1$ and $A_{44}L=B_4$ when $K=0$.
Therefore we obtain $\mathcal{L}(K=0)|\Psi\rangle\rangle=0$, meaning that $|\Psi\rangle\rangle$ is the steady state with $\lambda_0=0$.

Next, we calculate the left-eigenvector $\langle\langle \Psi'|$ with
\begin{align}
\mathcal{L}^\dagger(K)|\Psi'\rangle\rangle=0.
\end{align}
Similarly, we obtain
\begin{align}
M^\dagger_{n,{\rm aux}}=\begin{pmatrix}
     0 & -ih_n & ih^*_n & 0\\
    -ih^*_n & -\frac{B_1+B_4}{2} & 0 & ih^*_n\\
     ih_n & 0 & -\frac{B_1+B_4}{2} & -ih_n\\
     0 &  ih_n & -ih^*_n& 0
     \end{pmatrix},\quad
M^\dagger_{n,{\rm aux}}|\psi'_n\rangle\rangle=0,\quad|\psi'_n\rangle\rangle=
\begin{pmatrix}
\psi'_{n,1}\\
\psi'_{n,2}\\
\psi'_{n,3}\\
\psi'_{n,4}
\end{pmatrix}=
\begin{pmatrix}
1\\
0\\
0\\
1
     \end{pmatrix}
~~\Rightarrow~~
\mathcal{L}_{\rm aux}^\dagger|\Psi\rangle\rangle=0,
\quad
|\Psi'\rangle\rangle=\frac{1}{\sqrt{L}}\begin{pmatrix}
|\psi'_1\rangle\rangle\\
|\psi'_2\rangle\rangle\\
...\\
|\psi'_L\rangle\rangle
\end{pmatrix};
\end{align}
\begin{align}
|\Phi'\rangle\rangle&=\mathcal{L}^\dagger(K)|\Psi'\rangle\rangle=\big[\mathcal{L}^\dagger(K)-\mathcal{L}^\dagger_{\rm aux}(K)\big]|\Psi'\rangle\rangle
=\frac{1}{\sqrt{L}}[(A^*_{11}L-B_1),0,0,(A^*_{14}L-B_4),(A^*_{11}L-B_1),0,0,(A^*_{14}L-B_4),...]^T=0
\end{align}
since $A^*_{11}L-B_1=0$ and $A^*_{44}L-B_4=0$ when $K=0$.
Thus the left-eigenvector of the steady state is given by $|\Psi'\rangle\rangle$.

Finally, for a small but nonzero $K=k'-k$,
The Liouvillian can be written as
\begin{align}
\mathcal{L}(K)=\mathcal{L}(0)+\Delta\mathcal{L},\qquad
\Delta \mathcal{L}={\rm diag}[\Delta M_1,\Delta M_2,...,\Delta M_L]+N_L\otimes \Delta M_0,
\end{align}
 \begin{equation}
    \Delta M_n(K)=\begin{pmatrix}
     0 & -K\sum_s sJ^*_s e^{isk_n}& 0 & 0\\
     K\sum_s sJ_s e^{-isk_n}& 0 & 0 & 0\\
     0 & 0 & 0 & -K\sum_s sJ^*_s e^{isk_n}\\
     0 & 0 & K\sum_s sJ_s e^{-isk_n} & 0
     \end{pmatrix},
    \qquad
    \Delta M_0(K)=\begin{pmatrix}
    iK\sum_s s\gamma_s^{a,a}/L&0&0&0\\
    0&0&0&0\\
    0&0&0&0\\
    0&0&0&iK\sum_s s\gamma_s^{b,b}/L
     \end{pmatrix}.
  \end{equation}
The first-order perturbation to the steady-state eigenvalue $\lambda_0=0$ satisfies 
\begin{align}
&\Delta \lambda_0 \propto \langle\langle \Psi'| \Delta\mathcal{L}|\Psi\rangle\rangle
=i\frac{K}{L}\sum_n\Bigg(\sum_s s\gamma_s^{a,a}
+ \sum_s s\gamma_s^{b,b}\Bigg)=iK\Bigg(\sum_s s\gamma_s^{a,a}
+ \sum_s s\gamma_s^{b,b}\Bigg).\label{eq:perturbed_asymmetric}
\end{align} 
The Liouvillian spectral winding number is thus given by
\begin{align}
W_0={\rm Sgn}[\sum_s s\gamma_s^{a,a}
+ \sum_s s\gamma_s^{b,b}],
\end{align}
which depends solely on the dissipation parameters.

\section{Beyond the single particle cases}

\subsection{Two hard-core bosons}

\begin{figure}[t]
      \centering
      \includegraphics[width=0.75\linewidth]{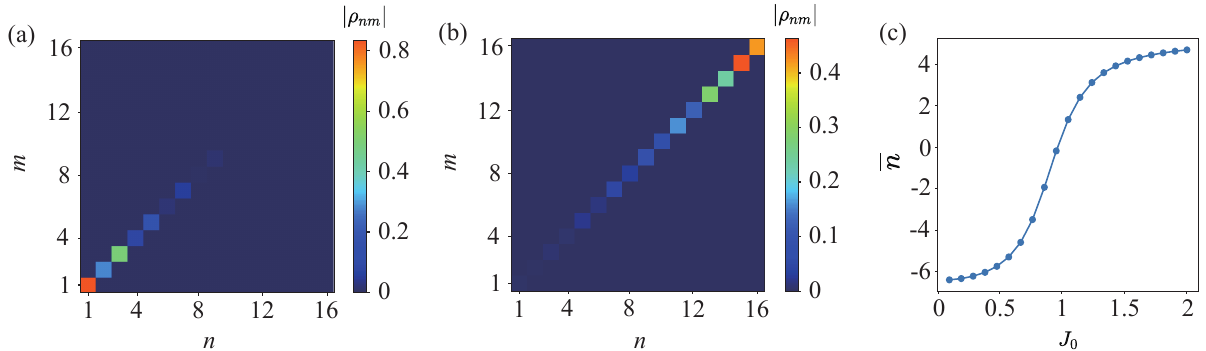}
      \caption{
    LSE of the model with two hard-core bosons. (a) and (b) reduced density matrix of the steady state with $J_0=0.5$ and $J_0=2$, respectively. (c) average position of
      the diagonal terms of density matrix $\overline{n}$ under different $J_0$. The other parameters are $J_1=1$, $\gamma_0=\gamma_1=1.2$, and $L=8$. }
      \label{fig:s5}
  \end{figure}

The creation operator for a hard-core boson is defined as $\hat{c}_{\alpha,l}^\dagger$, which acts on the vacuum state satisfying:
  \begin{equation}
    \hat{c}_{\alpha,l}^\dagger|0\rangle=|\alpha,l\rangle,  
  \end{equation}
  with the additional constraint prohibiting double occupancy:
  \begin{equation}
    (\hat{c}_{\alpha,l}^\dagger)^2=0.  
  \end{equation}
  Here, we consider the following Hamiltonian in terms of creation and annihilation operators,
  \begin{equation}
    \hat{H}^{(1)}=\sum_lJ_0\hat{c}_{a,l}^\dagger\hat{c}_{b,l}+J_1\hat{c}_{a,l+1}^\dagger\hat{c}_{b,l}+{\rm h.c.},   
  \end{equation}
  with jump operators in Eq.~(\ref{eq:jumpopsS}) given by
  \begin{equation}
    \hat{D}_{l,1}^{a,b}=\sqrt\gamma_1\hat{c}_{a,l+1}^\dagger\hat{c}_{b,l},~~ \hat{D}_{l,0}^{a,b}=\sqrt\gamma_0\hat{c}_{a,l}^\dagger\hat{c}_{b,l}.  
    \label{eq:twobody_dis}
  \end{equation}

  Based on the description above, we can write the Hamiltonian for two hard-core bosons as:
  \begin{equation}
    \hat{H}^{(2)}=\hat{P}(\hat{H}^{(1)}\otimes I+I\otimes\hat{H}^{(1)})\hat{P},  
  \end{equation}
  where
  \begin{equation}
    \hat{P}=\sum_{(\alpha,l)<(\alpha',l')}\hat{c}_{\alpha,l}^\dagger\hat{c}_{\alpha',l'}^\dagger|0\rangle\langle0|\hat{c}_{\alpha,l}\hat{c}_{\alpha',l'}
  \end{equation}
  is the projection operator.
  Similarly, the dissipative operator can be written as:
  \begin{equation}
    \hat{D}^{\alpha,\alpha'(2)}_{l,p}=\hat{P}(\hat{D}^{\alpha,\alpha'}_{l,p}\otimes I+I\otimes\hat{D}^{\alpha,\alpha'}_{l,p})\hat{P} 
  \end{equation}
  with $p=0,1$ in Eq.~\eqref{eq:twobody_dis}.
  Here, we vectorize the density matrix in the double space of bra and ket,
  and the Liouvillian superoperator can be written as:
  \begin{equation}
   \mathcal{L}=  -i(I\otimes \hat{H}^{(2)}-\hat{H}^{(2)T}\otimes I)  
   +\sum\left(\hat{D}^{\alpha,\alpha'(2)*}_{\alpha,l}\otimes\hat{D}^{\alpha,\alpha'(2)}_{\alpha,l}-\frac{1}{2}I\otimes \hat{D}^{\alpha,\alpha'(2)\dagger}_{\alpha,l} \hat{D}^{\alpha,\alpha'(2)}_{\alpha,l}-\frac{1}{2}(\hat{D}^{\alpha,\alpha'(2)\dagger}_{\alpha,l} \hat{D}^{\alpha,\alpha'(2)}_{\alpha,l})^T\otimes I \right). 
   \label{eq:L_2body}
 \end{equation}
  Diagonalizing Eq.~(\ref{eq:L_2body}) yields the steady-state density matrix $\rho^{(2)}$, which is defined on the two-body hard-core boson Hilbert space. To characterize the skin effect distribution in real space more clearly, we further compute the single-particle reduced density matrix:
  \begin{equation}
    \rho^{(1)}_{(\alpha,l),(\alpha',l')}={\rm Tr}\left(\rho^{(2)}\hat{c}_{\alpha',l'}^\dagger\hat{c}_{\alpha,l}\right). 
  \end{equation}
  Numerical results for different parameters are presented in Fig.\ref{fig:s5}. As seen in Fig.\ref{fig:s5}(a) and (b), when the Hamiltonian is in different topological phases, the LSE exhibits opposite directions, similar to the single-particle case. This is further illustrated by the phase diagram in Fig.\ref{fig:s5}(c).

\subsection{Two interacting bosons}
\begin{figure}
    \centering
    \includegraphics[width=1\linewidth]{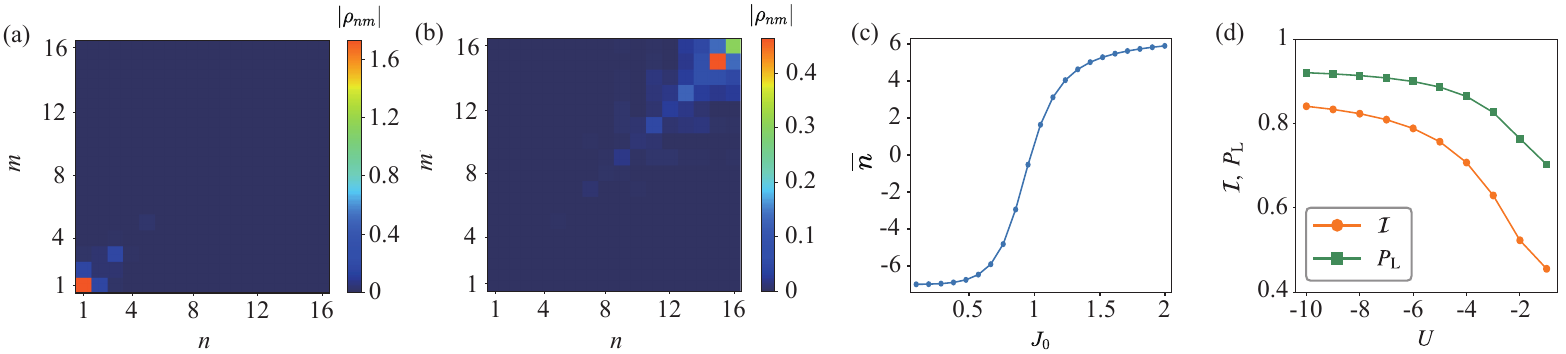}
    \caption{
    LSE of the model with two interacting bosons.
    (a) and (b) reduced density matrix of the steady state with $J_0=0.5$ and $J_0=2$, respectively. 
    (c) average position of the diagonal terms of density matrix $\overline{n}$ under different $J_0$. 
    The interaction strength in (a), (b), and (c) is $U=-5$.
    The inverse participation ratio ($\mathcal{I}$) and the left boundary occupation probability $P_L$ as functions of the interaction strength $U$.
    The other parameters are $J_1=1$, $\gamma_0=\gamma_1=1.2$, and $L=8$. }
    \label{fig:s6}
\end{figure}

To investigate the effect of interactions, we further consider a system of conventional bosons with onsite interaction, described by the Hamiltonian
\begin{equation}
\hat{H}=\sum_l\left(J_0 \hat{c}_{a,l}^{\dagger}\hat{c}_{b,l}
+J_1 \hat{c}_{a,l+1}^{\dagger}\hat{c}_{b,l}+\mathrm{h.c.}\right)
+\frac{U}{2}\sum_{l,\alpha=a,b}\hat{n}_{\alpha,l}\left(\hat{n}_{\alpha,l}-1\right)
\end{equation}
with $\hat{n}_{\alpha,l}=\hat{c}_{\alpha,l}^{\dagger}\hat{c}_{\alpha,l}$,
and the same quantum jumps as in Eq.~\eqref{eq:twobody_dis}.
Following the same procedure as in the previous subsection, we diagonalizing the Liouvillian superoperator for the interacting bosons and obtain the reduced density matrix.

Numerical results for different parameters are presented in Fig.\ref{fig:s6}. As shown in Figs.\ref{fig:s6}(a) and \ref{fig:s6}(b), 
the steady state exhibits opposite LSE in different topological phases of the Hamiltonian, similar to the case of hard-core bosons.
This is further illustrated by the phase diagram in Fig.\ref{fig:s6}(c).

To unveil the effect of interactions on the localization properties of the system, we define the inverse participation ratio $\mathcal{I}$ and the left boundary occupation probability $P_L$ as follows:
\begin{equation}
   \mathcal{I}=\frac{(\sum_n \rho_{nn}^{(1)})^2}{({\rm Tr}\rho^{(1)})^2},
   ~~~~~~P_L=\frac{\langle\hat{n}_{a,1}\rangle+\langle\hat{n}_{b,1}\rangle}{{\rm Tr}\rho^{(1)}},
\end{equation}
where $\rho^{(1)}$ is the reduced density matrix.
Fig.\ref{fig:s6}(d) shows the variation of these two quantities with interaction strength. It can be seen that a stronger attractive interaction enhances the localization at the boundary. 

\bibliography{references}
\end{document}